\documentstyle[12pt,preprint,aps,epsfig]{revtex}
\tighten
\newcommand{\beq}{\begin{eqnarray}}
\newcommand{\eeq}{\end{eqnarray}}

\newcommand{\bsga}{ b \to s \gamma}

\newcommand{\bsb}{ \bar{B_s}}
\newcommand{\as}{ \alpha_s }
\newcommand{\aem}{ \alpha_{em} }
\newcommand{\mw}{ M_W }
\newcommand{\mhp}{ M_{H^{+}}}
\newcommand{\nceff}{ N_c^{{\rm eff}} }

\def\etap{\eta^{\prime}}
\def\etapp{\eta^{(')}}
\def\acp{{\cal A}_{CP}}

\def\nce{N_c^{\rm eff}}
\def\ov{\overline}

\def\lsim{ {\ \lower-1.2pt\vbox{\hbox{\rlap{$<$}\lower5pt\vbox{\hbox{$\sim$}}}}\ } }
\def\gsim{ {\ \lower-1.2pt\vbox{\hbox{\rlap{$>$}\lower5pt\vbox{\hbox{$\sim$}}}}\ } }
\newcommand{\tab}[1]{Table \ref{#1}}
\newcommand{\fig}[1]{Fig.\ref{#1}}
\newcommand{\non}{\nonumber\\ }
\newcommand{\obar}[1]{\shortstack{{\tiny (\rule[.4ex]{1em}{.1mm})}\\ [-.7ex] $#1$}}

\title{\bf{ Branching ratios and CP-violating asymmetries of $B_s \to h_1 h_2 $
decays in the general two-Higgs doublet models} }
\author{ Dong Zhang, Zhenjun Xiao
\thanks{Email address: zxiao@ibm320h.phy.pku.edu.cn}, and Chong Sheng Li
\thanks{Email address: csli@ibm320h.phy.pku.edu.cn}\\
{\small Department of Physics, Peking University, Beijing,
 100871, People's Republic of China } }
\date{\today}
\begin{document}
\maketitle
\begin{abstract}
Based on the low-energy effective Hamiltonian with the generalized
factorization, we calculate the new physics contributions to branching
ratios and CP-violating asymmetries of the charmless hadronic decays
$B_s \to h_1 h_2$ in the standard model and the general two-Higgs doublet
models (models I, II, and III). Within the considered paramter space, we find the
following. (a) In models I and II, the new physics corrections are
always small in size and will be masked by other larger known theoretical
uncertainties. (b)  In model III, the new physics corrections to
the branching ratios of those QCD penguin-dominated decays
$\ov B_s \to K^0\etapp, K^+ K^{-*}, etc.$, are large in size and insensitive
to the variations of $\mhp$ and $\nceff$.
For tree- or  electroweak penguin-dominated decay modes, however,
the new physics corrections are very small in size.
(c) For $\ov B_s \to K^+ K^{-*}$ and other seven decay modes, the branching
ratios are at the level of $(1-3)\times 10^{-5}$ and will be measurable
at the future hadron colliders with large $b$ production.
(d) Among the studied thirty nine $B_s$  meson decay
modes, seven of them can have a CP-violating asymmetry ${\cal A}_{CP}$
larger than $20\%$ in magnitude. The new physics corrections are
small or moderate in magnitude.
(e) Because of its large and $\nceff$ stable branching ratio and CP violating
asymmetry,  the decay $\ov B_s \to K^+ K^{-*}$ seems to be the ``best" channel
to find CP violation of $B_s$ system through studies of two-body charmless
decays of $B_s$ meson.
\end{abstract}

\vspace{.5cm}

\newpage
\section{ Introduction } \label{sec:1}

In B experiments, new physics beyond the standard model (SM) may manifest itself,
for example,  in following two ways\cite{slac504,fj99}: (a) decays which are
expected to be rare in the SM are found to have large branching ratios; (b)
CP-violating asymmetries which are expected to vanish or be very small in the
SM are found to be significantly large or with a very different pattern with
what predicted in the SM. These potential deviations may be induced  by the
virtual effects of new physics through loop diagrams.

The observation of many two-body charmless hadronic $B_{u,d}$ meson decays by
CLEO, BaBar and Belle \cite{cleo99,cleo9912,cleo2001,babar2000,belle2000},
the successful start of the asymmetric B factories at SLAC and KEK, and the
expectation for large number of events of $B_{u,d}$ meson decays to be
accumulated at B
factories and other hadron colliders stimulated the intensive investigation
for various B decay channels. The two-body charmless hadronic decays
$B_{u,d} \to h_1 h_2$ [ where $h_1$ and $h_2$ are the light pseudo-scalar (P) and/or
vector(V) mesons ]  have been studied, for example, in Refs.
\cite{bh1h2,du97,ali98,ali9804,chenbs99,chen99}.

It is well known that the low energy effective Hamiltonian is the basic tool
to calculate the branching ratios and $\acp$ of B meson decays. The
short-distance QCD corrected Lagrangian at NLO level is available now
\cite{buchalla96,buras98},
but we do not know how to calculate hadronic matrix element from first
principles. One conventionally resort to the factorization approximation
\cite{bsw87}. However, we also know that non-factorizable contribution
really exists and can not be neglected numerically for most hadronic B
decay channels. To remedy the naive factorization hypothesis, some authors
\cite{cheng98,ali98,ali9804} introduced a
phenomenological parameter $\nceff$ (i.e. the effective number of color)
to model the non-factorizable contribution to hadronic matrix element,
which is commonly called the generalized factorization. Very recently, Cheng
{\it et al.} \cite{cheng99a} studied and resolved the controversies on the
gauge dependence and infrared singularity of the effective Wilson coefficients
$C_i^{eff}$ \cite{bs98} by using the perturbative QCD factorization
theorem.

Unlike the $B_{u,d}$ meson, the heavier $B_s$ meson can not be produced
in CESR, KEKB and PEP-II $e^+ e^-$ colliders. Only upper limits on decay
rates of several charmless hadronic $B_s$ decays are current available
from LEP collaborations \cite{pdg2000,l395}, such as $B_s \to K^+ K^-$,
$K^+ \pi^-$, $\pi^0 \eta$ and $B_s \to \eta \eta$, while most
of them are far beyond the theoretical predictions. However, it is expected
that many $B_s$ decays can be seen at the future hadron colliders with large
$b$ production. Recent theoretical studies and experimental measurements
about the mixing of $B_s^0-\bar{B}^0_s$ can be found in
Refs.\cite{beneke99,delphi00}. The early studies of two-dody charmless
hadronic decays of $B_s$ meson can be found in Refs.\cite{dbgn93,du1993}.
Based on the framework of generalized factorization,
Tseng\cite{tseng99} analyzed the exclusive charmless $B_s$ decays involving $\etapp$,
while Chen, Cheng and Tseng \cite{chenbs99} calculated the branching ratios
of thirty nine charmless two-body decays of $B_s$ meson.
It is found that the branching ratios of $\eta \etapp$
and several other decay modes can be as large as $10^{-5}$ and
measurable at future experiments.

In a recent work\cite{xiao2002}, we made a systematic study for the new physics
contributions to the branching ratios of seventy six $B_{u,d} \to h_1 h_2$ decay
channels in the framework of the general two-Higgs-doublet models
(2HDM's). In this paper  we extend the work to the case of $B_s$ meson.
In additional to the branching ratios, we here also calculate the new physics
contributions to the CP-violating asymmetries $A_{CP}$ of charmless hadronic
decays $B_s\to h_1 h_2$ induced by the new gluonic and eletroweak
charged-Higgs penguin diagrams in  the general 2HDM's ( models I, II and III).
Using the effective Hamiltonian with improved generalized factorization
\cite{cheng99a}, we  evaluate
analytically all new strong and electroweak penguin diagrams induced by
exchanges of charged Higgs bosons in the quark level processes
$b \to q V^*$ with $q \in \{d,s \}$ and $V\in \{ gluon,\; \gamma, Z\} $,
and then combine the new physics contributions with their SM counterparts and
finally calculate the branching ratios and CP-violating asymmetries for all
thirty nine exclusive $B_s\to h_1 h_2$ decay modes.

This paper is organized as follows. In Sec. \ref{sec:2hdm}, we describe the basic
structures  of the 2HDM's and examine the allowed parameter space of the general
2HDM's from  currently available data. In Sec. \ref{sec:heff}, we evaluate
analytically the new penguin diagrams and find the effective Wilson coefficients
$C_i^{eff}$ with the inclusion of new physics contributions, and present the
formulae needed to calculate the branching ratios ${\cal B} (B \to h_1 h_2)$.
In Sec. \ref{sec:bh1h2} and \ref{sec:acp}, we calculate and show numerical
results of branching  ratios and CP-violating asymmetries for thirty nine
$B_s$ decay modes, respectively. We focus on those decay modes with large
branching ratios and large CP-violating asymmetries. The conclusions and
discussions are included in the final section.

\section{ The general 2HDM's and experimental constraints } \label{sec:2hdm}

The simplest extension of the SM is the so-called two-Higgs-doublet
models\cite{2hdm}. In such models, the tree level flavor changing
neutral currents(FCNC's) are absent if one introduces an discrete
symmetry to constrain the 2HDM scalar potential and
Yukawa Lagrangian. Lets consider a Yukawa Lagrangian
of the form\cite{atwood97}
\beq
{\cal L}_Y &=&
\eta^U_{ij}\bar{Q}_{i,L} \tilde{\phi_1}U_{j,R} +
\eta^D_{ij}\bar{Q}_{i,L} \phi_1 D_{j,R}
+\xi^U_{ij}\bar{Q}_{i,L} \tilde{\phi_2}U_{j,R}
+\xi^D_{ij}\bar{Q}_{i,L} \phi_2 D_{j,R}+ h.c., \label{leff}
\eeq
where $\phi_{i}$ ($i=1,2$) are the two Higgs doublets of a
two-Higgs-doublet model, $\tilde{\phi}_{1,2}=
i\tau_2 \phi^*_{1,2}$, $Q_{i,L}$ ($U_{j,R}$) with $i=(1,2,3)$ are the
left-handed isodoublet quarks (right-handed   up-type quarks),
$D_{j,R}$  are the right-handed  isosinglet  down-type quarks,
while $\eta^{U,D}_{i,j}$  and $\xi^{U,D}_{i,j}$ ($i,j=1,2,3$ are
family index ) are generally the non-diagonal matrices of the  Yukawa
coupling. By imposing the discrete symmetry: $\phi_1 \to - \phi_1$, $
\phi_2 \to \phi_2$, $D_i \to - D_i$, and $U_i \to  \mp U_i$, one obtains
the so called model I and model II.

During past years, models I and II have been studied extensively in
literature and tested experimentally, and the model II has been very
popular since it is the building block of the minimal supersymmetric
standard model. In this paper, we focus on  the third
type of the two-Higgs-doublet model\cite{hou92}, usually known as the model III
\cite{atwood97,hou92}. In model III, no discrete symmetry is
imposed and both up- and down-type quarks then may have diagonal
and/or flavor changing couplings with $\phi_1$ and $\phi_2$.
As described in \cite{atwood97}, one can choose a suitable
basis $(H^0, H^1, H^2, H^\pm)$ to express two Higgs doublets.
The $H^\pm$ are the physical charged Higgs
boson, $H^0$ and $h^0$ are the physical CP-even neutral Higgs boson
and the $A^0$ is the physical CP-odd neutral Higgs boson. After the
rotation of quark fields, the Yukawa Lagrangian of quarks are of the
form \cite{atwood97},
\beq
{\cal L}_Y^{III} =
\eta^U_{ij}\bar{Q}_{i,L} \tilde{\phi_1}U_{j,R} +
\eta^D_{ij}\bar{Q}_{i,L} \phi_1 D_{j,R}
+\hat{\xi}^U_{ij}\bar{Q}_{i,L} \tilde{\phi_2}U_{j,R}
+\hat{\xi}^D_{ij}\bar{Q}_{i,L} \phi_2 D_{j,R} + H.c.,
\label{lag3}
\eeq
where $\eta^{U,D}_{ij}$ correspond to the diagonal mass matrices of
up- and down-type quarks, while the neutral and charged flavor changing
couplings will be \cite{atwood97}. We make the same ansatz on the $\xi^{U,D}_{ij}$
couplings as the Ref.\cite{atwood97}
\beq
\xi^{U,D}_{ij}=\frac{\sqrt{m_im_j}}{v} \lambda_{ij}, \ \
\hat{\xi}^{U,D}_{neutral}= \xi^{U,D}, \ \
\hat{\xi}^{U}_{charged}= \xi^{U}V_{CKM}, \ \
\hat{\xi}^{D}_{charged}= V_{CKM} \xi^{D}, \label{cxiud}
\eeq
where $V_{CKM}$ is the Cabibbo-Kobayashi-Maskawa mixing matrix
\cite{ckm}, $i,j=(1,2,3)$ are the generation index. The coupling
constants $\lambda_{ij}$ are free parameters to be determined by
experiments, and they may also be complex.

In model II and setting $1 \leq \tan{\beta}=v_2/v_1 \leq 50$ favored by
experimental measurements \cite{pdg2000}, the constraint on the mass
of charged Higgs boson due to CLEO data of $\bsga$ is $\mhp \geq 200$
GeV at the NLO level \cite{nlo2hdm}. For model I, however, the limit can
be much weaker due to the possible destructive interference with the SM amplitude.
For model III, the situation is not as clear as model II because there
are more free parameters here \cite{atwood97,aliev99}.
In a recent paper \cite{chao99}, Chao {\it et al.} studied the decay $\bsga$ by
assuming that only the couplings $\lambda_{tt}=|\lambda_{tt}| e^{i\theta_t}$
and $\lambda_{bb}=|\lambda_{bb}|e^{i\theta_b}$ are
non-zero. They found that the constraint on $\mhp$ imposed by the CLEO
data of $\bsga$ can be greatly relaxed by considering the phase
effects of $\lambda_{tt}$ and $\lambda_{bb}$. From the studies of
Refs.\cite{chao99,xiaonc}, we know that for model III the parameter space
\beq
&& \lambda_{ij}=0, \ \ for \ \ ij\neq tt,\ \ or \ \  bb, \nonumber\\
&& |\lambda_{tt}|= 0.3,\ \ |\lambda_{bb}|=35,\ \
\theta=(0^\circ - 30^\circ),\ \ \mhp=(200 \pm 100 ){\rm GeV}, \label{eq:lm3}
\eeq
are allowed by the available data, where $\theta=\theta_{bb}-\theta_{tt}$.

From the CERN $e^+ e^-$ collider (LEP) and the Fermilab Tevatron
searches for charged Higgs bosons \cite{gross99}, the new combined
constraint in the $(\mhp \tan{\beta})$ plane has been given, for
example, in Ref. \cite{pdg2000}: the direct lower limit is
$\mhp > 77 $ GeV, while $0.5 \leq \tan{\beta} \leq 60$ for a relatively
light charged Higgs boson with $\mhp \sim 100$ GeV.
Combining the direct and indirect limits together, we here conservatively
consider the range of $100 {\rm GeV } \leq \mhp \leq 300$ GeV, while take
$\mhp=200$ GeV as the typical value for models  I, II, and III. For models I
and II we consider the range of $1 \leq \tan{\beta} \leq 50$, while take
$\tan{\beta}=2$ as the typical value.

\section{Effective Hamiltonian in the SM and 2HDM's }  \label{sec:heff}

The standard theoretical frame to calculate the inclusive three-body decays
$b \to s \bar{q} q $\footnote{For $b \to d \bar{q} q$ decays, one simply make the
replacement $s \to d$.} is based on the effective Hamiltonian
\cite{buras98,ali9804,chen99},
\beq
{\cal H}_{eff}(\Delta B=1) = \frac{G_F}{\sqrt{2}} \left \{
\sum_{j=1}^2 C_j \left ( V_{ub}V_{us}^* Q_j^u  + V_{cb}V_{cs}^* Q_j^c \right )
- V_{tb}V_{ts}^* \left [ \sum_{j=3}^{10}  C_j Q_j  + C_{g} Q_{g} \right ] \right \}
\label{heff2}
\eeq
Here the first ten operators $Q_1 - Q_{10}$ can be found for example in
Refs.\cite{ali9804,chen99,xiao2002}, while the chromo-magnetic operator reads:
\beq
Q_{g}&=& \frac{g_s}{8\pi^2}m_b \bar{s}_\alpha \sigma^{\mu \nu}
(1+ \gamma_5)T^a_{\alpha \beta} b_{\beta} G^a_{\mu \nu}
\label{q8}
\eeq
where $\alpha$ and $\beta$ are the $SU(3)$ color indices, $T^a_{\alpha \beta}$
( $a=1,...,8$) are the Gell-Mann matrices. Following Ref.\cite{chenbs99},
we do not consider the effect of the weak annihilation and exchange diagrams.

The coefficients $C_{i}$ in Eq.(\ref{heff2}) are the well-known Wilson
coefficient. Within the SM and at scale $\mw$, the Wilson coefficients $C_1(M_W),
\cdots, C_{10}(\mw)$ and $C_{g}(\mw)$ have been given for example in
Refs.\cite{buchalla96,buras98}. By using QCD renormalization group equations, it is
straightforward to run Wilson coefficients $C_i(\mw)$ from the scale $\mu =0( \mw)$
down to the lower scale $\mu =O(m_b)$. Working consistently  to the NLO precision,
the Wilson coefficients $C_i$ for $i=1,\ldots,10$ are needed in NLO precision,
while it is sufficient to use the leading logarithmic value for $C_{g}$.

\subsection{New strong and electroweak penguins}

For the charmless hadronic decays of B meson under consideration, the new physics
will manifest itself by modifying the corresponding Inami-Lim functions
$C_0(x), D_0(x), E_0(x)$ and $E'_0(x)$ which determine the coefficients
$C_3(\mw), \ldots, C_{10}(\mw)$ and $C_{g}(\mw)$. These modifications, in turn,
will change the SM predictions of the branching ratios and CP-violating asymmetries
for decays $B_s \to h_1 h_2$ under study.

The new strong and electroweak penguin diagrams can be obtained from the
corresponding penguin diagrams in the SM by replacing the internal $W^{\pm}$
lines with the charged-Higgs $H^+$ lines.
In Ref.\cite{xiao2002}, we calculated analytically the new $Z^0$-,
$\gamma$- and gluon-penguin diagrams induced by the exchanges
of charged-Higgs boson $H^+$, and found the new $C_0, D_0, E_0$, and
$E'_0$ functions which describe the new physics contributions to the
Wilson coefficients through the new penguin diagrams,
\beq
C_0^{III} &=& \frac{-x_t}{16} \left[ \frac{y_t}{1-y_t} + \frac{y_t}{(1-y_t)^2}\ln[y_t]
\right ]\cdot |\lambda_{tt}|^2 \label{eq:c0m3}\\
D_0^{III} &=& -\frac{1}{3} H(y_t)|\lambda_{tt}|^2 \label{eq:d0m3}\\
E_0^{III} &=& -\frac{1}{2}I(y_t)|\lambda_{tt}|^2,
\label{eq:e0m3}\\
{E'_0}^{III}&=& \frac{1}{6}J(y_t)|\lambda_{tt}|^2
 - K(y_t) |\lambda_{tt} \lambda_{bb}| e^{i\theta},\label{eq:e0pm3}
\eeq
with
\beq
H(y)&=& \frac{38 y - 79 y^2 + 47 y^3}{72(1-y)^3}
+ \frac{4y -6y^2 + 3y^4}{12 (1-y)^4} \ln[y]\label{eq:hhy}\\
I(y)&=& \frac{16y -29y^2 +7 y^3}{36(1-y)^3} + \frac{2y- 3y^2}{6(1-y)^4}\log[y],
\label{eq:iiy} \\
J(y)&=& \frac{2y + 5y^2 - y^3}{4(1-y)^3} + \frac{3y^2}{2(1-y)^4}\log[y],
\label{eq:jjy} \\
K(y)&=& \frac{-3y + y^2}{4(1-y)^2} - \frac{y}{2(1-y)^3}\log[y].
\label{eq:kky}
\eeq
where $x_t=m_t^2/\mw^2$, $y_t=m_t^2/\mhp^2$, and the small terms proportional to
$m_b^2/m_t^2$ have been neglected. In models I and II,  one can find the corresponding
functions $C_0$, $D_0$,$E_0$ and $E'_0$ by evaluating the new strong and electroweak
penguins in the same way as that in model III:
\beq
C_0^{I} &=& C_0^{II} = \frac{-x_t}{8 \tan^2{\beta}} \left[ \frac{y_t}{1-y_t}
    + \frac{y_t}{(1-y_t)^2}\ln[y_t] \right ], \label{eq:c0m2} \\
D_0^{I} &=& D_0^{II} =  -\frac{2}{3\tan^2{\beta}} H(y_t), \label{eq:d0m2}\\
E_0^{I} &=& E_0^{II} -\frac{1}{\tan^2{\beta}}I(y_t), \label{eq:e0m2}\\
{E'_0}^{I}&=& \frac{1}{3 \tan^2{\beta}}\left [ J(y_t) - 6 K(y_t)\right],
\label{eq:e0pm1}\\
{E'_0}^{II}&=& \frac{1}{3 \tan^2{\beta}}J(y_t) + 2 K(y_t),\label{eq:e0pm2}
\eeq
where $y_t=m_t^2/\mhp^2$, $\tan{\beta}=v_2/v_1$ where $v_1$ and $v_2$ are the
vacuum expectation values of the Higgs doublet $\phi_1$ and $\phi_2$
as defined before.

Combining the SM part and the new physics part together, the NLO Wilson
coefficients $C_i(\mw)$ and $C_g(\mw)$ can be written as
\beq
C_1(\mw) &=& 1 - \frac{11}{6} \; \frac{\as(\mw)}{4\pi}
               - \frac{35}{18} \; \frac{\aem}{4\pi} \, , \label{eq:c1mw}\\
C_2(\mw) &=&     \frac{11}{2} \; \frac{\as(\mw)}{4\pi} \, , \\
C_3(\mw) &=& -\frac{\as(\mw)}{24\pi} \left [ E_0(x_t) +E_0^{NP} -\frac{2}{3}
\right ]\non
&& +\frac{\aem}{6\pi} \frac{1}{\sin^2\theta_W}
             \left[ 2 B_0(x_t) + C_0(x_t) + C_0^{NP} \right] \, ,\\
C_4(\mw) &=& \frac{\as(\mw)}{8\pi} \left [ E_0(x_t)+E_0^{NP}
-\frac{2}{3} \right ] \, ,\\
C_5(\mw) &=& -\frac{\as(\mw)}{24\pi}
\left [E_0(x_t)+E_0^{NP} -\frac{2}{3} \right ] \, , \\
C_6(\mw) &=& \frac{\as(\mw)}{8\pi}
\left [E_0(x_t)+E_0^{NP} -\frac{2}{3} \right ] \, ,\\
C_7(\mw) &=& \frac{\aem}{6\pi} \left [ 4 C_0(x_t) + 4 C_0^{NP}
    + D_0(x_t) +D_0^{NP} -\frac{4}{9}\right ]\, , \\
C_8(\mw) &=& C_{10}(\mw)= 0 \, , \\
C_9(\mw) &=& \frac{\aem}{6\pi} \left\{ 4C_0(x_t) + 4 C_0^{NP}
    +D_0(x_t) +D_0^{NP} -\frac{4}{9}\right. \non
&& \left. +  \frac{1}{\sin^2\theta_W} \left [ 10 B_0(x_t)
- 4 C_0(x_t) + 4 C_0^{NP} \right ]  \right\} \, , \label{eq:cimw}\\
C_{g}(\mw) &=& -\frac{1}{2}\left ( E'_0(x_t) + {E'}_0^{NP} \right )
\, ,\label{eq:c8gmw}
\eeq
where $x_t=m_t^2/M_W^2$, the functions $B_0(x)$, $C_0(x)$, $D_0(x)$, $E_0(x)$
and $E'_0$ are the familiar Inami-Lim functions \cite{inami81} in the SM
and can be found easily, for example, in Refs. \cite{buchalla96,epj983}.

Since the heavy new particles appeared in the 2HDM's have been integrated out
at the scale $\mw$, the QCD running of the Wilson coefficients $C_i(\mw)$ down
to the scale $\mu=O(m_b)$ after including the new physics
contributions will be the same as in the SM:
\beq
{\bf C}(\mu)&=&U(\mu, \mw) {\bf C}(\mw), \label{eq:cmu}\\
C_g(\mu)&=& \eta^{14/23}C_g(\mw) + \sum_{i=1}^{8} \bar{h}_i
\eta^{a_i}, \label{eq:cgmu}
\eeq
where ${\bf C}(\mw)=( C_1(\mw), \ldots, C_{10}(\mw))^{T}$, $U(\mu,
\mw)$ is the five-flavor $10\times 10$ evolution matrix at NLO level
as defined in Ref.\cite{buchalla96}, $\eta=\alpha_s(\mw)/\alpha_s(\mu)$,
and the constants $\bar{h}_i$ and $a_i$ can also be found in
Ref.\cite{buchalla96}.

In the NDR scheme and for $SU(3)_C$, the effective Wilson coefficients
\footnote{In the improved generalized factorization approach \cite{cheng99a}, these
effective coefficients are renormalization scale- and scheme-independent, gauge
invariant and infrared safe.} can be written as \cite{chen99}
\beq
C_i^{eff} &=& \left [ 1 + \frac{\alpha_s}{4\pi} \, \left( \hat{r}_V^T +
 \gamma_{V}^T \log \frac{m_b}{\mu}\right) \right ]_{ij} \, C_j
 +\frac{\alpha_s}{24\pi} \, A_i' \left (C_t + C_p + C_g \right)
+ \frac{\alpha_{ew}}{8\pi}\, B_i' C_e ~, \label{eq:wceff}
\eeq
where $A_i'=(0,0,-1,3,-1,3,0,0,0,0)^T$, $B_i'=(0,0,0,0,0,0,1,0,1,0)^T$, the
matrices  $\hat{r}_V$ and $\gamma_V$ contain the process-independent
contributions from the vertex diagrams. The matrix
$\gamma_V$ and $\hat{r}_V$ have been given explicitly, for example, in Eq.(2.17) and (2.18) of
Ref.\cite{chen99}. Note that the correct value of the element $(\hat{r}_{NDR})_{66}$ and
$(\hat{r}_{NDR})_{88}$ should be  17 instead of 1 as pointed in Ref.\cite{cheng00a}.

The function $C_t$, $C_p$, and $C_g$ describe the contributions arising
from the penguin diagrams of the current-current
$Q_{1,2}$ and the QCD operators $Q_3$-$Q_6$, and the tree-level diagram of the
magnetic dipole operator $Q_{8G}$, respectively. We here  also follow the
procedure of Ref.\cite{ali98} to include the contribution of magnetic gluon
penguin. The functions $C_t$, $C_p$, and $C_g$ are given in  the NDR scheme by
\cite{ali9804,chen99}
\beq
C_t &=& \left [ \frac{2}{3} + {\lambda_u\over\lambda_t}G(m_u)
    +{\lambda_c\over\lambda_t} G(m_c) \right ] C_1,\label{cct} \\
C_p &=& \left [\frac{4}{3} - G(m_q) - G(m_b) \right ] C_3
    + \left [ \frac{10}{3} - \sum_{i=u,d,s,c,b} G(m_i) \right ] (C_4+C_6),  \label{ccp}\\
C_e &=& {8\over9}\left [ \frac{2}{3} +  {\lambda_u\over\lambda_t} G(m_u)
    + {\lambda_c\over\lambda_t} G(m_c) \right ] (C_1+3C_2),\label{cce}\\
C_g &=& -{2m_b\over \sqrt{< k^2>}}C^{\rm eff}_{g}, \label{ccg}
\eeq
with $\lambda_{q'}\equiv V_{q'b}V_{q'q}^*$, and $C_g^{eff} = C_{g}(\mu) + C_5$.
The function $G(m)$ can be found, for example, in Refs. \cite{chen99,xiao2002}.
For the two-body exclusive
B meson decays any information on $k^2$ is lost in the factorization assumption,
one usually  use the "physical" range for $k^2$ \cite{ali9804,chenbs99,chen99}:
$\frac{m_b^2}{4}\stackrel{<}{\sim} k^2 \stackrel{<}{\sim}\frac{m_b^2}{2}$.
Following Refs.\cite{ali9804,chenbs99,chen99} we take $k^2=m_b^2/2$ in the numerical
calculation.

\subsection{Decay amplitudes in the BSW model} \label{sec:bsw}

Following Ref.\cite{chenbs99}, the possible effects of final state
interaction (FSI) and the contributions
from annihilation channels will be neglected although they may play a
significant rule for some decay modes.
The new physics effects on the B decays under study will be included by using
the modified effective coefficients $a_i$ ($i=3,\dots,10$) as given in the
second entries of \tab{ai:bd} and \tab{ai:bs} for the model III. The
effective coefficients $a_i$ in models I and II are not shown explicitly in
\tab{ai:bd} and \tab{ai:bs}. In the numerical calculations
the input parameters as given in Appendix and Eq.(\ref{eq:lm3}) will
be used implicitly.

With the factorization ansatz \cite{bsw87}, the three-hadron
matrix elements or the decay amplitude $<XY|H_{eff}|B_s>$ can be factorized
into a sum of products of two current matrix elements $<X|J_1^\mu|0>$ and
$<Y|J_{2\mu}|B_s>$ ( or $<Y|J_1^\mu|0>$ and $<X|J_{2\mu}|B_s>$).
The explicit expressions of matrix elements can be found, for example,
in Refs.\cite{bsw87,bijnens92}.

In the B rest frame, the branching ratios of two-body B meson decays can
be written as
\beq
{\cal  B}(B_s \to X Y )=  \tau_{B_s}\,  \frac{|p|}{8\pi M_{B_s}^2}
|M(B_s \to XY)|^2\label{eq:brbpp}
\eeq
for $B_s \to P P$ decays, and
\beq
{\cal  B}(B_s \to X Y )=   \tau_{B_s} \, \frac{|p|^3}{8\pi M_V^2}
|M(B_s \to X Y )/(\epsilon \cdot p_{B})|^2\label{eq:brbpv}
\eeq
for $B_s \to P V$ decays. Here $\tau(B_s^0)=1.493 ps$\cite{pdg2000}, $p_B$ is the
four-momentum of the B meson, $M_V$ and $\epsilon$ is the mass and polarization
vector of the produced light vector meson respectively,  and $|p|$
is the magnitude of momentum of particle X and Y in the B rest frame,
\beq
|p| =\frac{1}{2M_B}\sqrt{[M_B^2 -(M_X + M_Y)^2] [ M_B^2 -(M_X-M_Y)^2 ]} \label{eq:pxy}
\eeq

For $B_s \to VV$ decays, the situation is more involved. One needs to evaluate
the helicity matrix element $H_\lambda = <V_1(\lambda) V_2(\lambda)|H_{eff}|B)>$
with $\lambda=0, \pm 1$. The branching ratio of the decay $B \to V_1 V_2$ is given
in terms of $H_\lambda$ by
\beq
{\cal  B}(B_s \to V_1 V_2 ) &=&  \tau_{B_s}\,
\frac{|p|}{8\pi M_B^2}\left ( |H_0|^2 + |H_{+1}|^2 + |H_{-1}|^2 \right )
\eeq
where $|p|$ has been given in Eq.(\ref{eq:pxy}). The three independent
helicity amplitudes $H_0$, $H_{+1}$ and $H_{-1}$ can be expressed by three
invariant amplitudes $a, b, c$ defined by the decomposition
\beq
H_\lambda &=& i\epsilon^\mu(\lambda)\eta^\nu(\lambda)\left[
    a g_{\mu\nu}+\frac{b}{M_1 M_2}p_\mu
    p_\nu + \frac{ic}{M_1 M_2}\epsilon_{\mu\nu\alpha\beta}p_1^\alpha
    p^\beta \right] \label{eq:hl}
\eeq
where $p_{1,2}$ and $M_{1,2}$ are the four momentum and masses of $V_{1,2}$,
respectively. $p=p_1 + p_2 $ is the four-momentum of B meson, and
\beq
H_{\pm 1} &=& a \pm c \sqrt{x^2-1}, ~~~~
H_0 = -ax - b\left ( x^2-1 \right ) \label{eq:h01} \\
x &=& \frac{M_B^2-M_1^2-M_2^2}{2M_1 M_2}
\eeq
For individual decay mode, the coefficients $a, b$ and $c$ can be determined
by comparing the helicity amplitude $H_\lambda = <V_1(\lambda) V_2(\lambda)
|H_{eff}|B_s>$ with the expression (\ref{eq:hl}).

In the generalized factorization approach, the effective Wilson coefficients
$C_i^{eff}$ will appear in the decay amplitudes in the combinations:
\beq
a_{2i-1}\equiv C_{2i-1}^{{\rm eff}} +\frac{{C}_{2i}^{{\rm eff}}}{\nceff}, \ \
a_{2i}\equiv C_{2i}^{{\rm eff}}     +\frac{{C}_{2i-1}^{{\rm eff}}}{\nceff}, \ \ \
(i=1,\ldots,5) \label{eq:ai}
\eeq
where the effective number of colors $\nceff$ is treated as a free parameter varying in
the range of $2 \leq \nceff \leq \infty$, in order to model the non-factorizable
contribution to the hadronic matrix elements.
Although  $\nceff$ can in principle vary from channel to channel,
but in the energetic two-body hadronic B
meson decays, it is expected to be process insensitive as supported by
the data \cite{chenbs99}.  As argued in ref.\cite{cheng98},
$\nceff(LL)$ induced by the $(V-A)(V-A)$ operators can be rather
different from $\nceff(LR)$ generated by  $(V-A)(V+A)$ operators.
Since we here focus on the calculation of new physics
effects on the studied B meson decays induced by the new
penguin diagrams in the two-Higgs doublet models, we will simply assume that
$\nceff(LL)\equiv \nceff(LR)=\nceff$ and consider the variation
of $\nceff$ in the range of $2 \leq \nceff \leq \infty$.
For more details about the cases of
$\nceff(LL)\neq \nceff(LR)$, one can see for example Ref.\cite{chenbs99}.
We here will not  consider the possible effects of final state
interaction (FSI) and the contributions from annihilation channels
although they may play a significant rule for some decay modes.

Using the input parameters as given in Appendix, and assuming $k^2=m_b^2/2$,
$\mhp=200$ GeV, the theoretical predictions of effective
coefficients $a_i$ are calculated and displayed in \tab{ai:bd} and \tab{ai:bs}
for the transitions $b\to d$ ( $\bar{b} \to \bar{d}$ ) and
$b\to s$ ($\bar{b} \to \bar{s}$), respectively.  For coefficients $a_3,
\ldots, a_{10}$, the first and second entries in tables
(\ref{ai:bd},\ref{ai:bs}) refer to the values of $a_i$ in the SM and model III,
respectively.

Compared with Ref.\cite{chenbs99}, the effective coefficients $a_i$ given
here have two new features:

\begin{itemize}
\item
The effective Wilson coefficients $C_i^{eff}$ here are not only
renormalization scale- and scheme-independent, but also gauge
invariant and infrared safe.

\item
The contribution due to the chromo-magnetic dipole operator $Q_g$ has been
included here through the function $C_g$ as given in Eq.(\ref{ccg}). For the
penguin dominated decay channels, operator $C_g$ will play an important
role.

\item
The coefficient $a_1$ and $a_2$ remain unchanged in 2HDM's since the new
physics considered here does not contribute through tree diagrams.

\item
The new physics contributions  are significant to coefficient $a_4$ and
$a_6$, but negligibly small  to coefficients $a_{3,5}$ and $a_{7-10}$.

\end{itemize}

All branching ratios here are the averages of the
branching ratios of $B$ and anti-$B$ decays. The ratio $\delta {\cal  B}$
describes the new physics correction on the decay ratio and is defined as
\beq
\delta {\cal  B} (B_s \to XY) = \frac{{\cal  B}(B_s \to XY)^{NP}
-{\cal  B}(B_s \to XY)^{SM}}{{\cal  B}(B_s \to XY)^{SM}} \label{eq:dbr}
\eeq

\section{Branching ratios of $B_s$ meson decays}\label{sec:bh1h2}

Using the formulas and input parameters as given in last section and in
Appendix,
it is straightforward to find the branching ratios for the thirty nine
$B_s \to PP, PV, VV $ decay channels. In the numerical
calculations, we use the decay amplitudes as given in Appendix A, B and C
of Ref.\cite{chenbs99} directly without further discussions about details.

Following Ref.\cite{bsw87,chenbs99}, the hadronic charmless $B$ meson decays
can be
classified into six classes: the first and last three classes correspond
to the tree-dominated  and penguin-dominated amplitudes, respectively.
\begin{itemize}
\item
Class-I and Class-II decays  are dominated by the external and internal
W-emission tree-diagrams, respectively. Examples are $\ov B_s \to
K^+ \pi^-, K^0 \pi^0, \cdots.$

\item
Class-III: the  decays involving both external and internal
W-emissions. But this class does not exist for the $B_s$ decays.

\item
Class-IV  and Class-V decay modes are governed by effective
coefficients $a_{4,6,8,10}$ and $a_{3,5,7,9}$, respectively.
Examples are $\ov B_s \to K^+ K^-, \pi \etapp, \cdots.$

\item
Class-VI decays involve the interference of class-IV and class-V
decays.
\end{itemize}

In tables \ref{brbssm}-\ref{brbsm2}, we present the numerical results of the
branching ratios for the thirty nine $B_s \to PP, PV , VV $ decays in
the framework of the SM and models I, II and III. Theoretical predictions
are made by using the central values of input parameters as given in
Eq.(\ref{eq:lm3}) and Appendix, and assuming $A=0.804$, $\lambda=0.22$,
$\rho=0.16$, $\eta=0.34$, $\mhp=200$GeV, $\theta=0^\circ,
30^\circ$, $\tan{\beta}=2$ and $\nceff=2, 3, \infty$ in  the generalized
factorization approach.
The $k^2$-dependence of the branching ratios is small in the range of
$k^2=m_b^2/2\pm 2\; GeV^2$ and hence the numerical results are given by fixing
$k^2=m_b^2/2$.

The SM predictions for all $B_s$ decay modes as listed in
tables \ref{brbssm} and \ref{brbslcr} are agree well with those given in
Ref.\cite{chenbs99}. The effect of changing $\hat{r}_V$ and including the
new contribution from
the chromo-magnetic operator $Q_g$ in the SM is not significant.

For decay modes involving $B_s \to K^*$ or $B_s \to \phi$ transitions, we use
two different set of form factors: the Bauer, Stech, and Wirbel (BSW)
form factor\cite{bsw87}  and the light-cone sum rule(LCSR) form factor
as given explicitly in Appendix. For decay modes
$B_s \to \pi^0 \phi, \phi \etap, \rho^0\phi, \omega \phi$ and $B_s \to \phi
\phi$, the variation of the branching ratios induced by using different set
of form factors is about a factor of 2, but small or moderate for all other
decay modes.

From numerical results, we see  the following general features of new
physics corrections:

\begin{itemize}
\item
In model III, the new physics corrections to QCD-penguin dominated decay
modes, such as $B_s \to K^0\etapp, \etapp \etapp, K^0 \bar{K}^0, etc.$,
are large in size and  insensitive to the variations of the mass
$\mhp$ and $\nceff$: from $~30\%$ to $~130\%$ $w.r.t$ the SM
predictions for both cases of $\theta=0^\circ, 30^\circ$. For tree-dominated
or  electroweak penguin dominated decay modes, however,  the new physics
corrections are very small in size: $\delta {\cal B} \leq 5\%$.

\item

In models I and II, the new physics corrections to all $B_s \to h_1 h_2$ decay
modes  are always small in size within the considered parameter space: less
than $10\%$ and $20\%$ in model I and II respectively, as shown in tables \ref{brbsm1}
and \ref{brbsm2}. So small corrections will be masked by other larger known
theoretical uncertainties. Variation of $\tan {\beta}$ in the range
of $2 \leq \tan{\beta} \leq 50 $ can not change this feature.

\item
In model III, the new gluonic penguins will contribute effectively through
the mixing of chromo-magnetic operator $Q_g$ with QCD penguin operators
$Q_3-Q_6$, as shown in Eq.(\ref{eq:wceff}). The $C_g^{eff}$ will
strongly dominate the new physics contributions to $B_s$ meson decays.
The branching ratios for all thirty nine decay modes have a very weak
dependence on $\theta$ in the range of $0^\circ \leq \theta \leq 30^\circ$.

\end{itemize}

As pointed in Refs.\cite{chenbs99,fleischer94}, the decays
\beq
\overline{B}_s \to \eta \pi, \etap \pi, \eta \rho, \etap\rho, \phi \pi,
\phi\rho \label{eq:small}
\eeq
do not receive any QCD penguin contributions, and are predominately governed
by $a_9$ and hence $\nceff$ insensitive. In 2HDM's, this remain to be true
because the new physics corrections to coefficients $a_{7-10}$ are negligibly
small as shown in tables \ref{ai:bd} and \ref{ai:bs}, and therefore, the new physics
contributions to these decay modes are also very small: $\leq 2\%$.
As suggested in Ref.\cite{chenbs99}, a measurement of these six decay modes can be
utilized to fix the parameter $a_9$. It is clear that the inclusion of new physics
contributions in the 2HDM's does not change this picture.

\begin{figure}[t] %fig.1
\vspace{-60pt}
\begin{minipage}[]{0.96\textwidth}
\centerline{\epsfxsize=\textwidth \epsffile{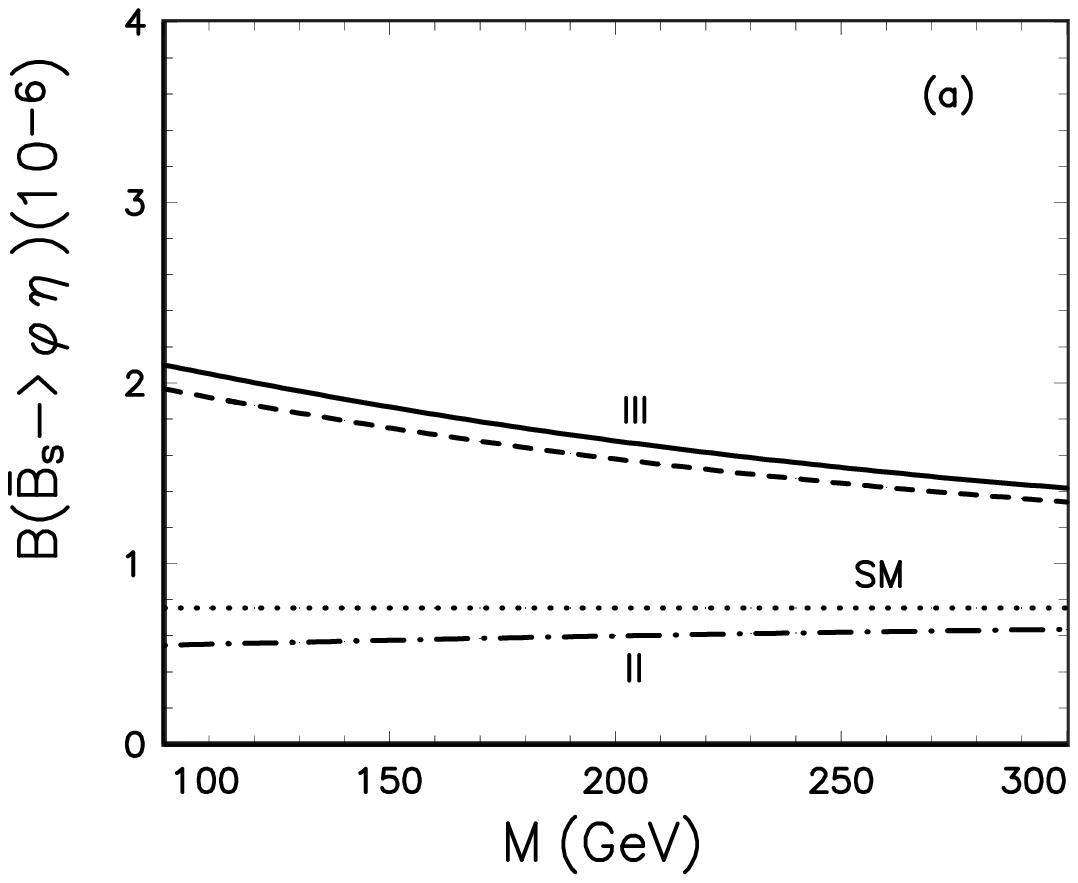}}
\vspace{-80pt}
\centerline{\epsfxsize=\textwidth \epsffile{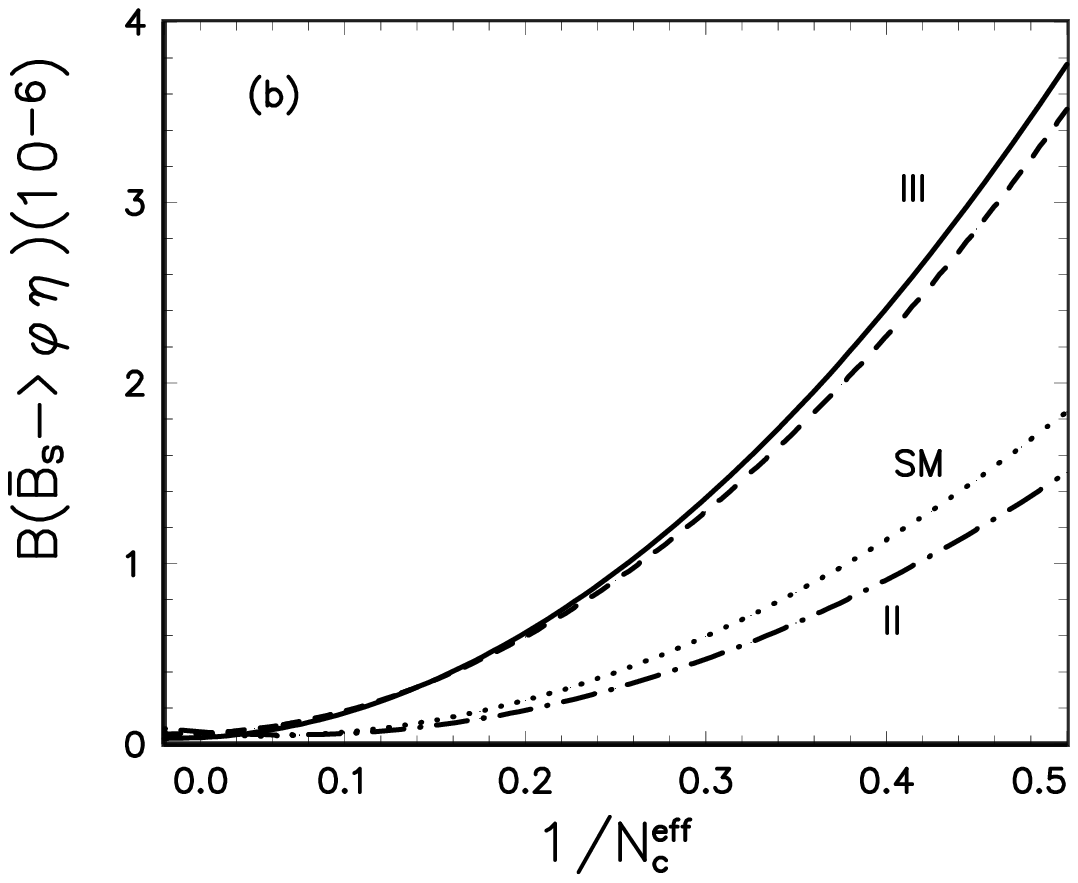}}
\vspace{-20pt}
\caption{Branching ratios ${\cal B}(\overline{B}_s \to \phi \eta)$ versus $\mhp$
and $1/\nceff$ in the SM and models II and III by using the BSW form factors.
For (a) and (b), we set $\nceff=3$ and
$\mhp=200$GeV, respectively. The four curves correspond to the theoretical predictions in the SM
(dotted line), model II (dot-dashed curve), model III with $\theta=0^\circ$ (solid curve) and
$\theta=30^\circ$ (short-dashed curve), respectively. }
\label{fig:fig1}
\end{minipage}
\end{figure}

For the decays
\beq
\overline{B}_s \to \omega \eta, \omega \etap, \phi \etapp, K\phi, K^*\phi, \phi \pi,
\eeq
the SM electroweak penguin corrections are in general as important as QCD penguin
effects, and very sensitive on $\nceff$. The new physics corrections to these
decay modes in the model III also have a strong dependence on the variation
of $\nceff$: $\delta {\cal B} =-20\% - 110\%$
for $2\leq \nceff \leq \infty$. As illustrated in Fig. \ref{fig:fig1},
for example, the branching ratio of $B_s \to \phi\eta$ decay has a
moderate $\mhp$-dependence, but a strong $\nceff$ dependence. For
Fig. \ref{fig:fig1}(a) and \ref{fig:fig1}(b), we set $\nceff=3$ and
$\mhp=200$ GeV, respectively. The four curves correspond to the theoretical
predictions in the SM (dotted curve), model II (dot-dashed curve),
model III with $\theta=0^\circ$
(solid curve) and $\theta=30^\circ$ (short-dashed curve), respectively.

\begin{figure}[t] %fig.2
\vspace{-60pt}
\begin{minipage}[]{0.96\textwidth}
\centerline{\epsfxsize=\textwidth \epsffile{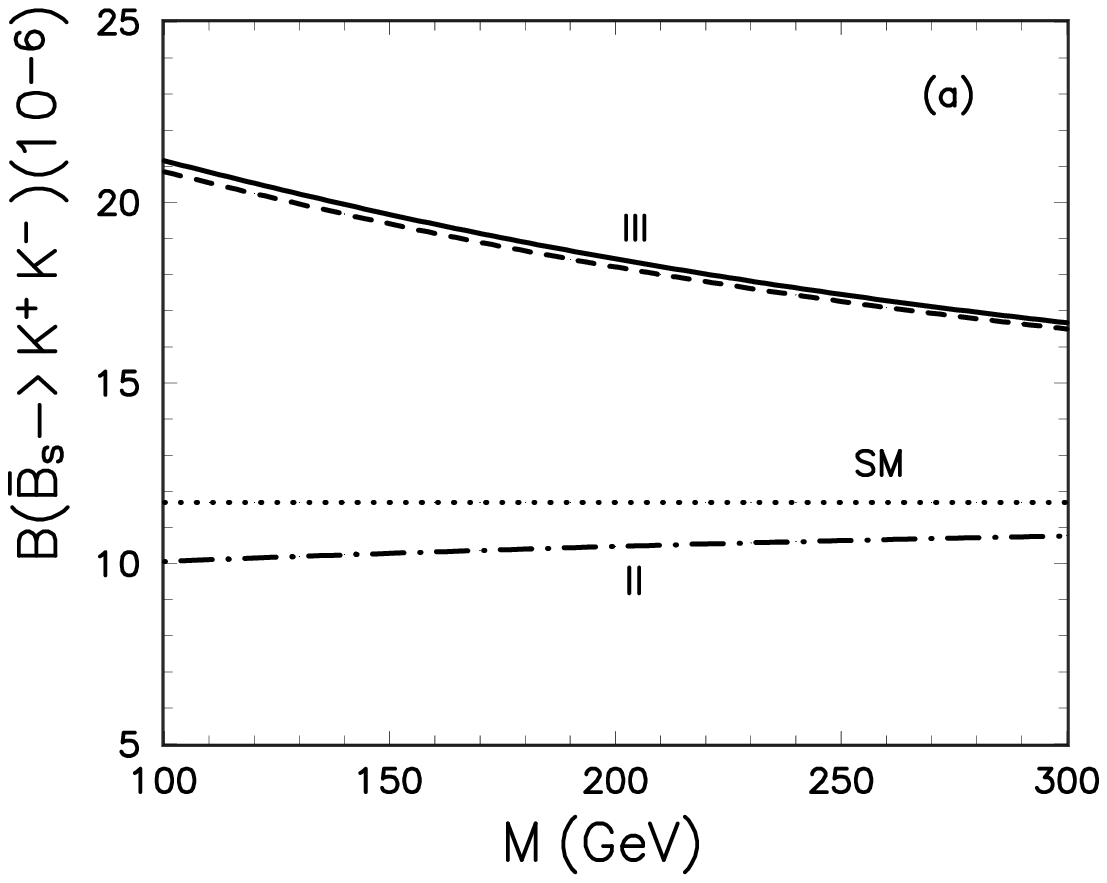}}
\vspace{-80pt}
\centerline{\epsfxsize=\textwidth \epsffile{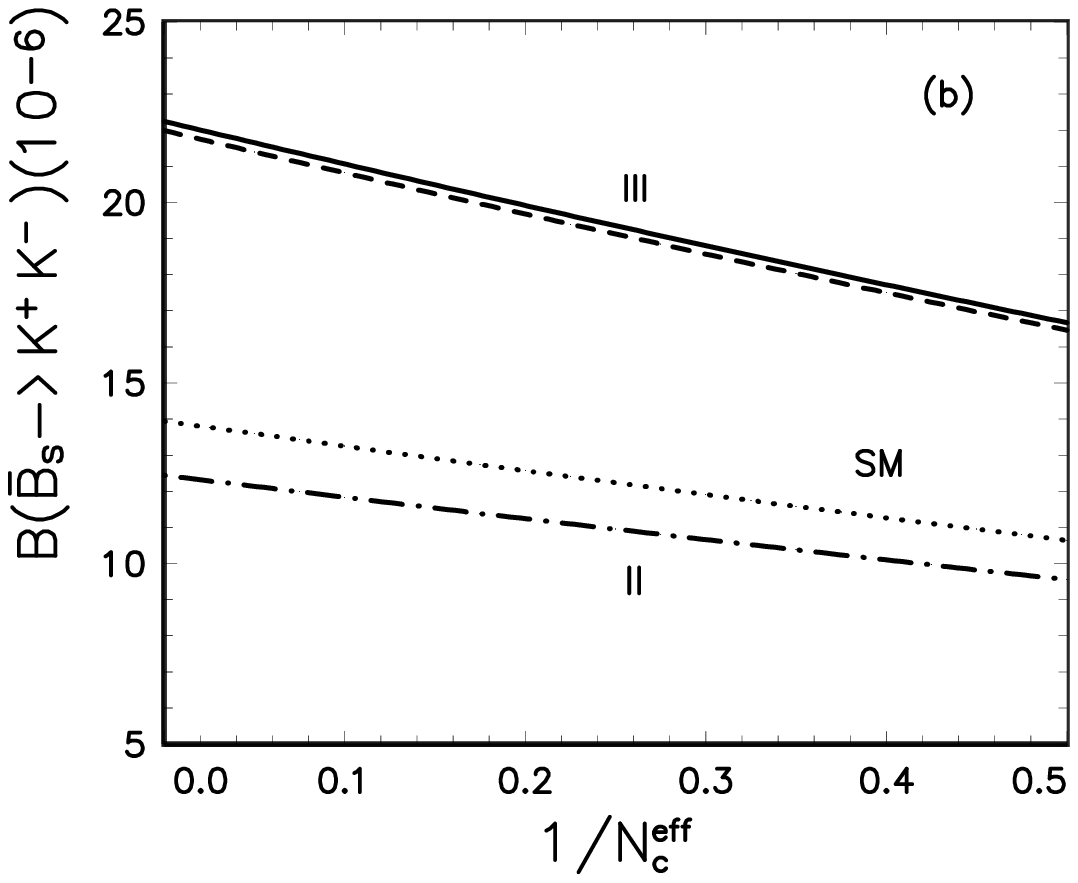}}
\vspace{-20pt}
\caption{Same as \fig{fig:fig1} but for the decay $\overline{B}_s \to K^+ K^- $. }
\label{fig:fig2}
\end{minipage}
\end{figure}

\begin{figure}[t] %fig.3
\vspace{-60pt}
\begin{minipage}[]{0.96\textwidth}
\centerline{\epsfxsize=\textwidth \epsffile{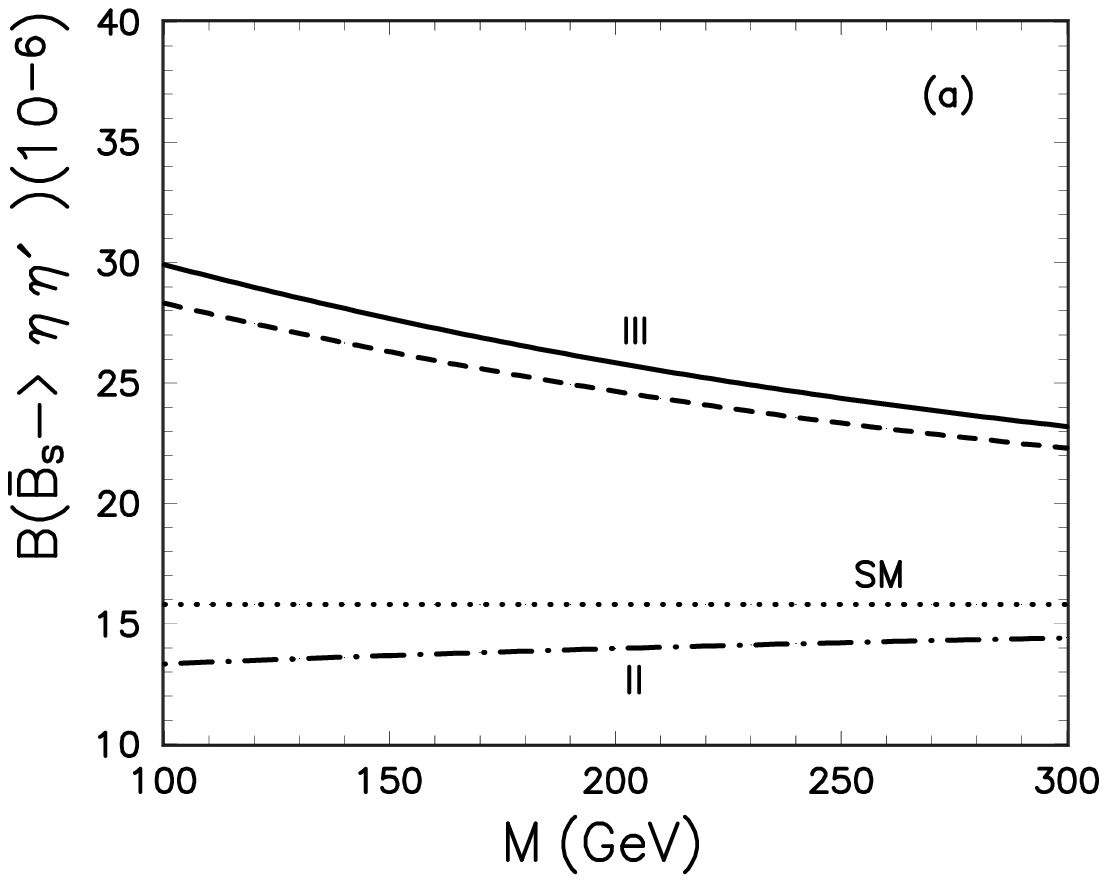}}
\vspace{-80pt}
\centerline{\epsfxsize=\textwidth \epsffile{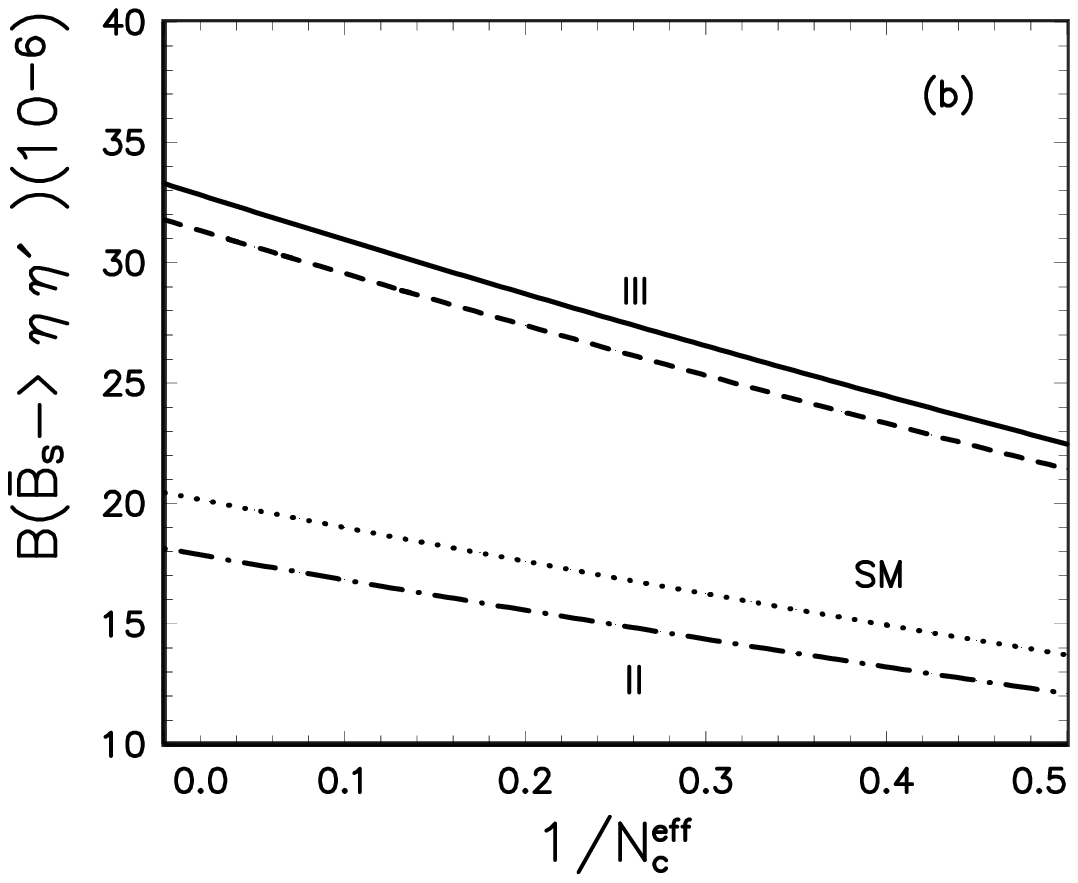}}
\vspace{-20pt}
\caption{Same as \fig{fig:fig1} but for the decay $\overline{B}_s \to \eta \etap$.}
\label{fig:fig3}
\end{minipage}
\end{figure}

Among the thirty nine charmless two-body hadronic $B_s$ decays,
we find that only 7 (8) of them have branching ratios at the level
of $10^{-5}$ in the SM (model III):
\beq
\overline{B}_s \to K^+ K^-, K^0 \overline{K}^0, \eta \eta, \etap \etapp,
K^+\rho, K^{+*} \rho^-, \phi\phi. \label{eq:large8}
\eeq
Among these eight decay modes, the new physics correction to the Class-I decay
mode $\overline{B}_s \to K^+\rho^-$ and $K^{+*} \rho^-$ are very small, from
$-2\%$to $1\%$. For remaining six decay modes, the new physics enhancement
are significant: from $\sim 50\%$ to $\sim 130\%$ and insensitive to the
variation of $\nceff$. These decay modes will be measurable at the future
hadron colliders with large $b$ production \cite{chenbs99}. In
Figs. \ref{fig:fig2} and \ref{fig:fig3}, we plot the mass and $\nceff$
dependence of the branching ratios of $\overline{B}_s \to K^+ K^-$ and $\eta
\etap$ decay modes.

After inclusion of new physics contributions in the models I, II and III, the
patterns observed in Ref.\cite{chenbs99} remain unchanged,
\beq
&& \Gamma(\ov B_s\to K^+K^-)
>\Gamma(\ov B_s\to K^+K^{*-})\gsim \Gamma(\ov B_s\to K^{*+}K^{*-})>
\Gamma(\ov B_s\to K^{+*}K^-),   \non
&& \Gamma(\ov B_s\to K^0\ov K^0)
>\Gamma(\ov B_s\to K^{0}\ov K^{*0})\gsim \Gamma(\ov B_s\to K^{*0}\ov K^{*0})>
\Gamma(\ov B_s\to K^{*0}\ov K^0).
\eeq

Recently, large decay rates for $B_{u}^+ \to K^+ \etap$ and $B_d \to K^0
\etap$ decays have been reported by CLEO and BaBar collaborations
\cite{cleo9912,cleo2001}.
The CLEO measurement of $B_d^0 \to K^0 \etap $ decay is
${\cal B}(B_d^0 \to K^0 \etap)= (89 ^{+18}_{-16}\pm 9)\times
10^{-6}$, which is larger than the branching ratios of $B \to K \pi$ decays
by a factor of 3 to 5.
For $B_s$ decays, the decay modes $\ov B_s \to \eta \etap$ and
$\ov B_s \etap \etap$ are the analogue of $B_d \to K^0 \etap$ decay,
and are expected to have large branching ratios.
From Table \ref{brbssm}, one can see that the SM predictions of the
branching ratios ${\cal B}(B_s \to \eta \etap )$ and  ${\cal B}(B_s
\to \etap \etap )$ are indeed large,
but in comparable size with other six decay modes listed in
Eq.(\ref{eq:large8}). The new physics enhancement to
these two decay modes are significant in size, $\sim 70\%$ in model
III, as illustrated in Fig. \ref{fig:fig3}.
After inclusion of new physics contributions, we find numerically that
\beq
&&{\cal B} (\ov B_s \to \eta \etap) \approx (23-33)\times 10^{-6}~, \\
&&{\cal B} (\ov B_s \to \etap \etap) \approx (12-16)\times 10^{-6}~.
\eeq
These theoretical predictions will be tested by the future experimental
measurements.

\begin{figure}[t] %fig.4
\vspace{-60pt}
\begin{minipage}[]{0.96\textwidth}
\centerline{\epsfxsize=\textwidth \epsffile{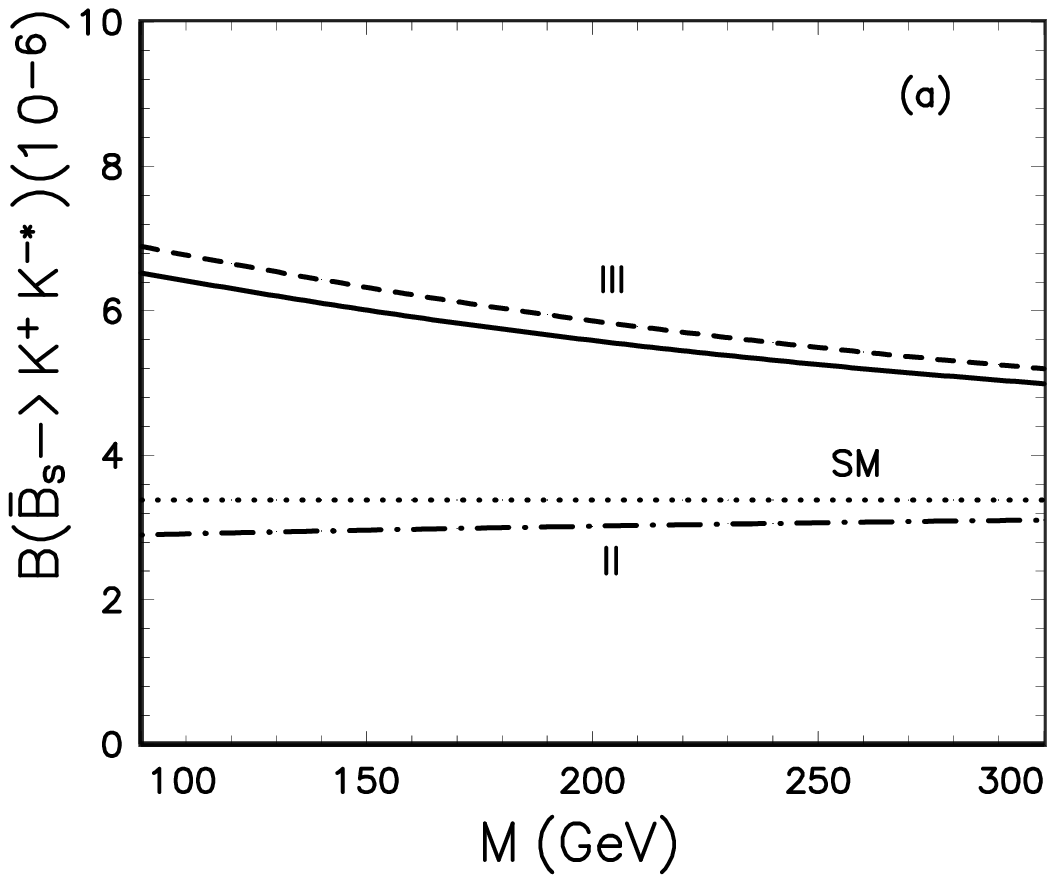}}
\vspace{-80pt}
\centerline{\epsfxsize=\textwidth \epsffile{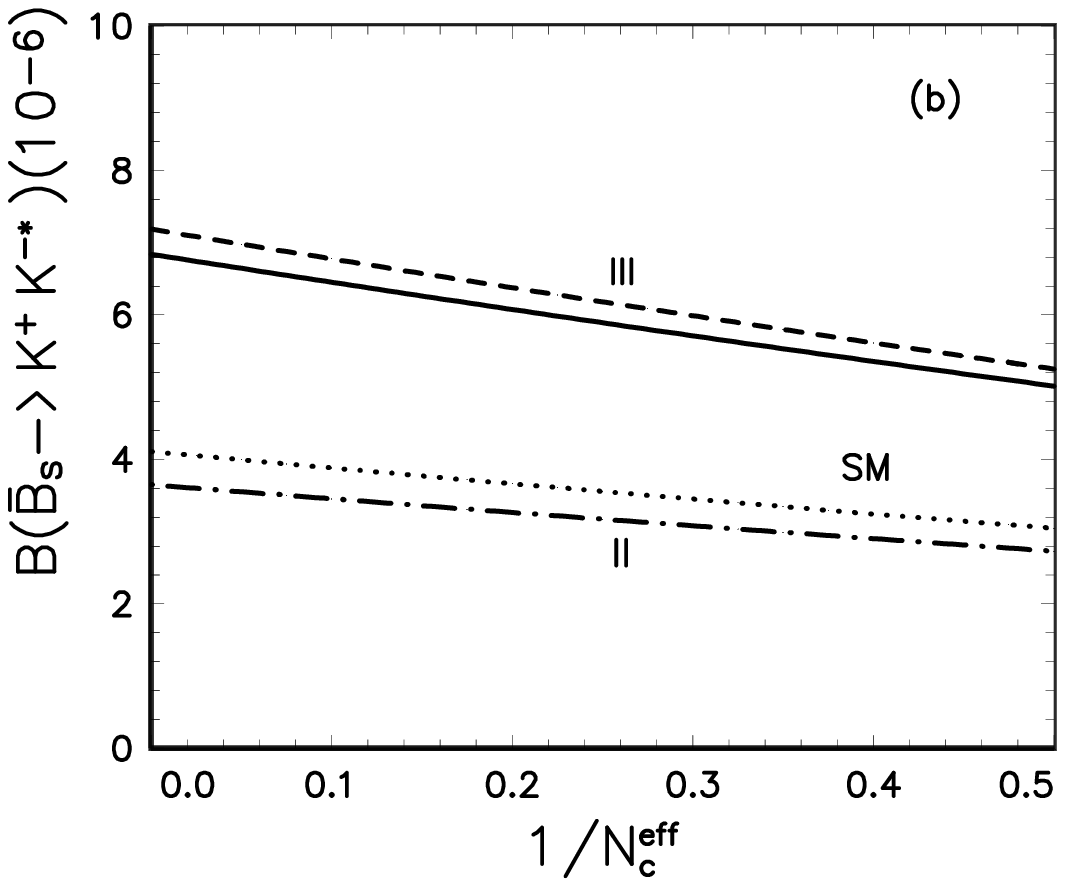}}
\vspace{-20pt}
\caption{Same as \fig{fig:fig1} but for the decay $\overline{B}_s \to K^+ K^{-*}$.}
\label{fig:fig4}
\end{minipage}
\end{figure}

\begin{figure}[t] %fig.5
\vspace{-60pt}
\begin{minipage}[]{0.96\textwidth}
\centerline{\epsfxsize=\textwidth \epsffile{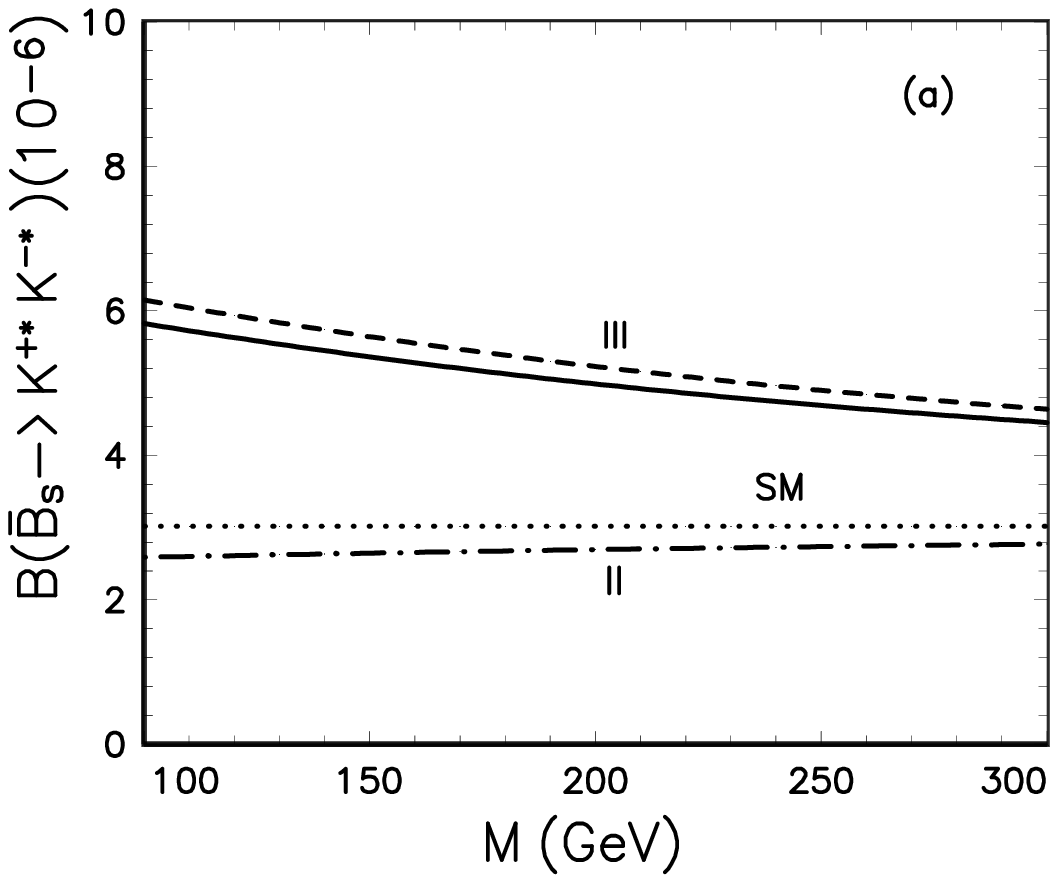}}
\vspace{-80pt}
\centerline{\epsfxsize=\textwidth \epsffile{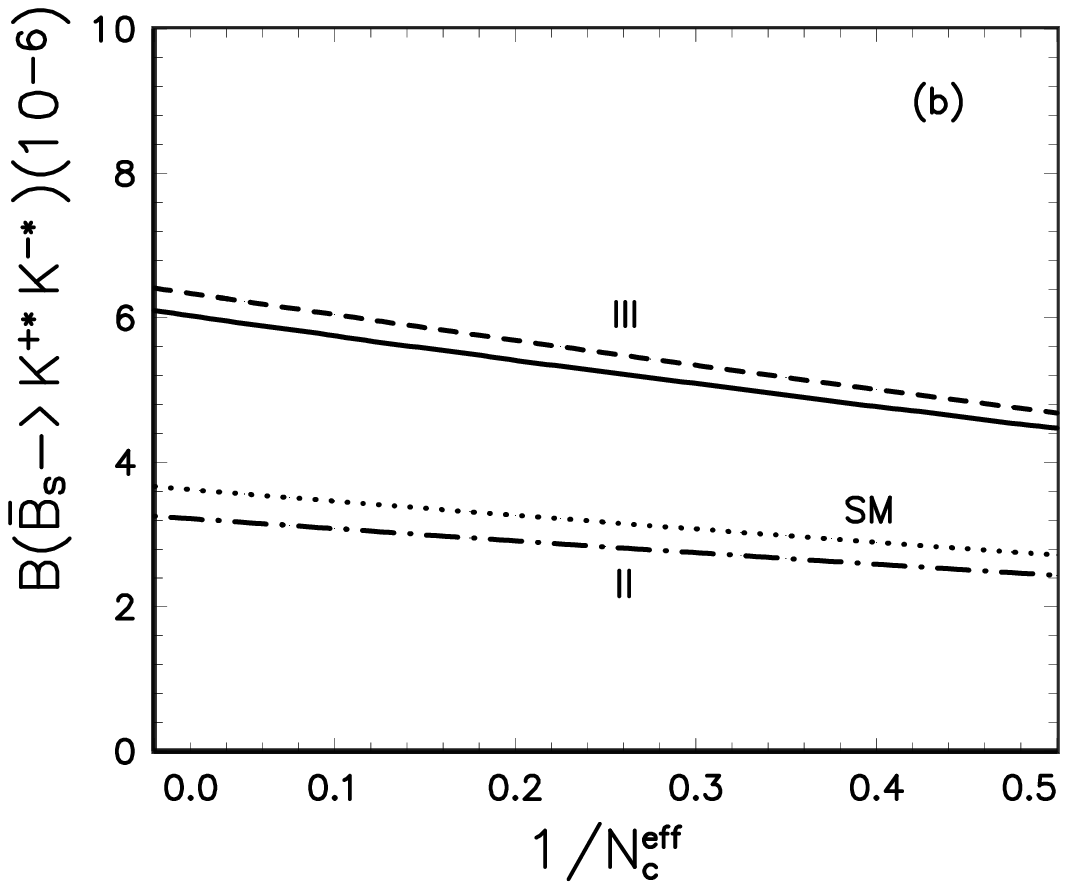}}
\vspace{-20pt}
\caption{Same as \fig{fig:fig1} but for the decay $\overline{B}_s \to K^{+*} K^{-*}$.}
\label{fig:fig5}
\end{minipage}
\end{figure}

For decays $\ov B\to K^+ K^{-*}$ and $\ov B\to K^{+*} K^{-*}$, they
have relatively large decay rates and weak $\mhp$ and $\nceff$ dependence.
In Figs. \ref{fig:fig4} and \ref{fig:fig5}, we plot the mass and $\nceff$
dependence of the branching ratios ${\cal B}(\ov B_s \to K^+ K^{-*})$
and ${\cal B}(K^{+*} K^{-*})$. It is easy to see that the new physics
contributions in the model III
to these two Class-IV decays are significant ( $\sim 70\%$ ) in size
and insensitive to the variations of $\mhp$ and $\nceff$.

\section{CP-violating asymmetries of $B_s$ meson decays}\label{sec:acp}

In Ref.\cite{du1993}, Du et al. studied the branching ratios and CP-violating
asymmetries for decay modes $B_s\to K^- \pi^+, K^+ K^-, \bar{K}^0 \pi^0,
\phi \phi$  and $\bar{K}^0 \phi$. Recently, Ali et al. \cite{ali9805}
estimated the CP-violating asymmetries in seventy six charmless hadronic
decays of $B_u$ and $B_d$ mesons. The calculation of the CP-violating
asymmetry $\acp$ for $B_s$ meson decays are theoretically very similar
with those of the $B_d$ meson decays. For more details about the theoretical
aspects of CP-violating asymmetries in $B_{u,d} \to h_1 h_2 $ decays,
one can see Ref.\cite{ali9805} and reference therein.  In this section,
we calculate the CP-violating asymmetries of $B_s \to h_1 h_2$ decays
in the framework of the SM and the general two-Higgs-doublet models.
We focus on evaluating the  new physics effects on $\acp$ for thirty nine
$B_s$ decay channels induced by charged Higgs penguin diagrams appeared
in the general two-Higgs-doublet models.

In models I and II, one does not expect sizable changes in $\acp$ of $B_s$
decays since
there is no any new phase introduced when compared with the SM. In model III,
although the introduce of a new phase $\theta $ played an important role in
relaxing the constraint on the parameter space of model III due to the CLEO
measurement of $B \to X_s \gamma$ decay as studied in Ref.\cite{chao99}, we
still do not expect dramatic changes for the pattern of the CP-violating
asymmetries of $B_s$ decays under consideration because this phase may alter
the theoretical prediction of $\acp$ through loop diagrams only.

Analogous to the $B_d$ meson decays,  the time dependent CP
asymmetry for the decays of states that were tagged as pure $B_s^{0}$ or
$\bar{B}_s^0$ at production is defined as
\beq
{\cal  A}_{CP}(t) = \frac{\Gamma(B_s^0(t) \to f ) -\Gamma(\bar{B}_s^0(t) \to \bar{f} )}{
\Gamma(B_s^0(t) \to f) + \Gamma(\bar{B}_s^0(t) \to \bar{f})}\label{eq:acp0}
\eeq

Following Ref.\cite{ali9805}, the neutral $B_s^0$ ($\bar{B}_s^0$) decays can
be classified into three classes according to the properties of the final
states $f$ and $\bar{f}$:
\begin{itemize}
\item
Class-1: $B_s^0 \to f$, $\bar{B}_s^0 \to \bar{f}$, and the final states $f$ or
$\bar{f}$  is not a common final state of $B_s^0$ and
$\bar{B}_s^0$, for example, $B_s^0 \to K^+ \pi^-$. The CP-violating asymmetry
for class-1 decays will be independent of time:
\beq
{\cal  A}_{CP} = \frac{\Gamma(B_s^0 \to f) -\Gamma(\overline{B}_s^0 \to \bar{f})}{
\Gamma(B_s^0 \to f) + \Gamma(\overline{B}_s^0 \to \bar{f})}\label{eq:acp1}
\eeq
in terms of partial decay widths.

\item
Class-2 and 3: $\obar{B_s^0} \to (f = \bar{f})$ with $f^{CP}=\pm f $ (class-2) or
$f^{CP}\neq \pm f $ (class-3), the time-integrated CP asymmetries are of the form
\beq
{\cal  A}_{CP}=\frac{1}{1 + x^2} \frac{1}{1 + |\lambda_{CP}|^2 }
- 2 \frac{ x }{ 1 + x^2 } \frac{ {\rm Im}(\lambda_{CP})}{ 1+ |\lambda_{CP}|^2}
\label{eq:acpti}
\eeq
with
\beq
\lambda_{CP}=\frac{V^*_{tb}V_{ts}}{V_{tb}V^*_{ts}}
\frac{ <f|H_{eff}|\bar{B}_s^0> }{ <f|H_{eff}|B_s^0>} \label{eq:lambda}
\eeq
where $x=\Delta M_{B_s^0}/\Gamma_{B_s^0} \approx 20$ is the preferred
value in the SM \cite{du1993} for the case of $B_s^0-\bar{B}_s^0$ mixing
\footnote{From Ref.\cite{pdg2000}, the upper limit is
$x=\Delta M_{B_s^0}/\Gamma_{B_s^0} > 15.7$ at $95\% C.L.$}.
Contrary to the $B_d$ meson decay where $x\approx 0.73$, it is easy to see
that the parameter $x$ for $B_s^0$ decays is very large. The first and
second term in Eq.(\ref{eq:acpti}) is strongly suppressed by $1/x^2$ and
$1/x$, respectively. We therefore do not expect large CP-violating
asymmetries $\acp$ for the class-2 and class-3 $B_s^0$ decays.
This expectation is  confirmed by the numerical results given below.

\end{itemize}

In  Tables \ref{acpsm1} and -\ref{acpm1}, we present  numerical results of
CP-violating asymmetries ${\cal  A}_{CP}$ for thirty nine
$B_s \to h_1 h_2$ decay channels in the SM and 2HDM's, using the
input parameters as given in Appendix, and assuming that
$k^2=m_b^2/2$, $\rho=0.16$, $\eta=0.34$, $\mhp=200$GeV, $\theta=0^\circ,
30^\circ$, and $\nceff=2,\; 3,\; \infty$.
We show the numerical results for the case of using BSW form factors only
since the differences induced by using the BSW or LCSR form factors
are small for almost all $B_s$ decay modes.

Among thirty nine $B_s$ decay modes studied, we find that seven of them
have CP-violating asymmetries larger than $ 20\%$ in the SM and model III:
\beq
\ov B_s \to K^{0*}\pi^0, K^0 \rho^0, \bar{K}^{0*} \etapp,
K^+ K^{-*}, K^{0*} \rho^0, K^{0*} \omega~.
\label{eq:acp10}
\eeq
All these seven decay modes are belong to the CP-class-1 decay modes.
On the other hand, all twenty four class-2 and 3 decay modes  have small
CP-violating asymmetries only, $|\acp| \lsim 5\%$, mainly due to the
strong suppression of $1/x^2$ as shown in Eq.(\ref{eq:acpti}).

\begin{figure}[t] %fig.6
\vspace{-60pt}
\begin{minipage}[]{0.96\textwidth}
\centerline{\epsfxsize=\textwidth \epsffile{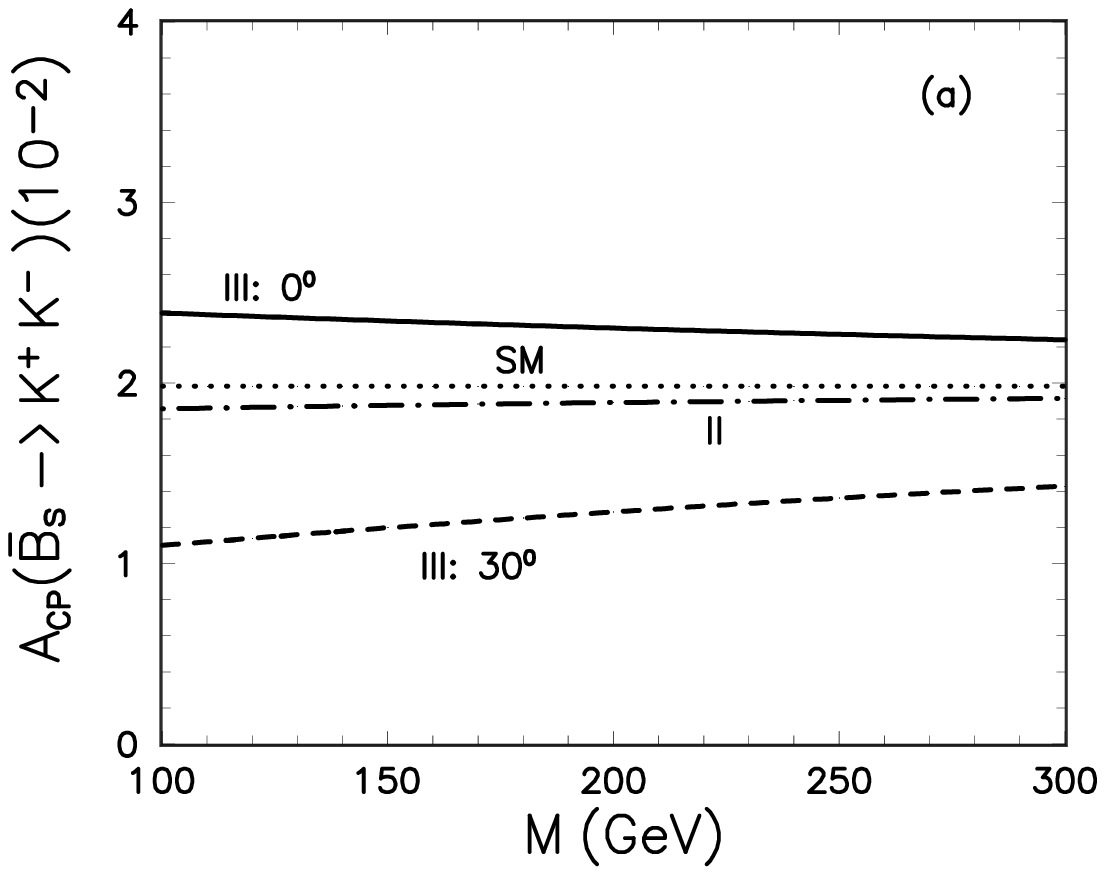}}
\vspace{-80pt}
\centerline{\epsfxsize=\textwidth \epsffile{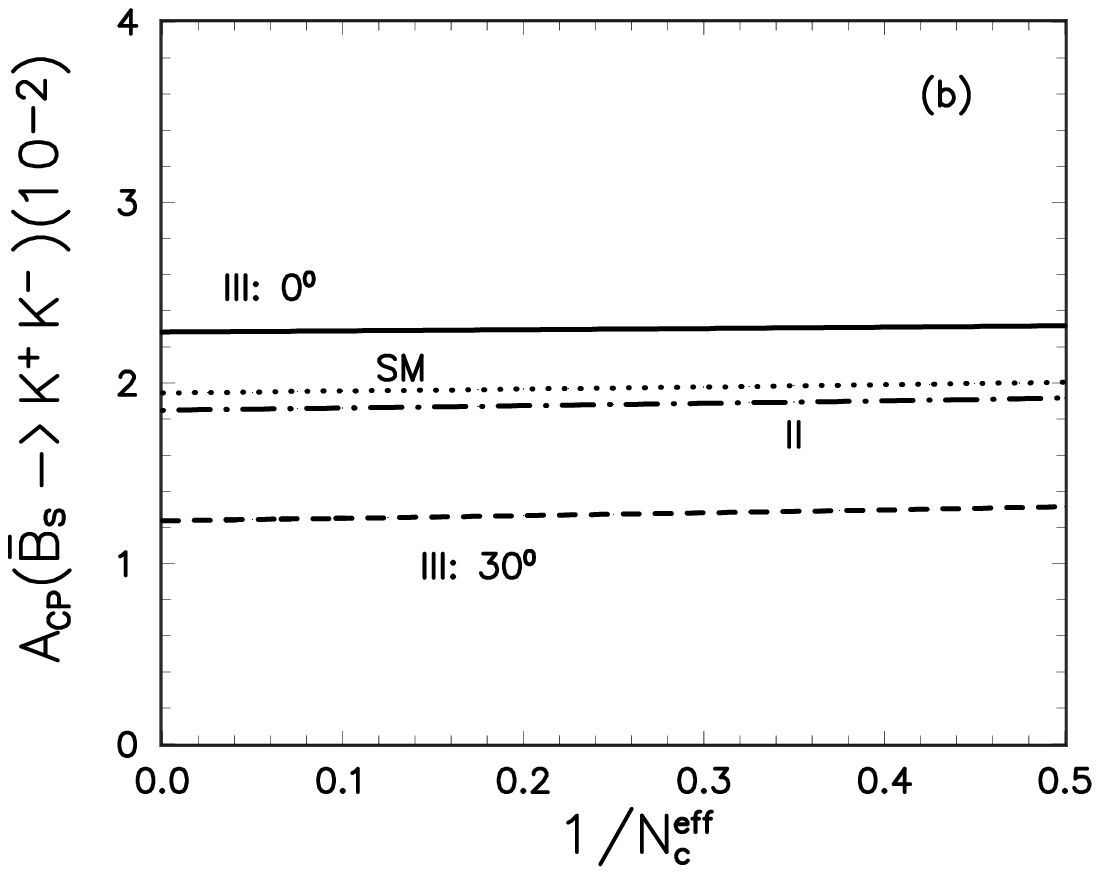}}
\vspace{-20pt}
\caption{CP-violating asymmetries $A_{CP}$ of  $\overline{B}_s \to K^+ K^-$ decay versus $\mhp$
and $1/\nceff$ in the SM and models II and III. For (a) and (b), we set $\nceff=3$ and
$\mhp=200$GeV, respectively. The four curves correspond to the theoretical predictions in the SM
(dotted line), model II (dot-dashed curve), model III with $\theta=0^\circ$ (solid curve) and
$\theta=30^\circ$ (short-dashed curve), respectively.}
\label{fig:fig6}
\end{minipage}
\end{figure}

\begin{figure}[t] %fig.7
\vspace{-60pt}
\begin{minipage}[]{0.96\textwidth}
\centerline{\epsfxsize=\textwidth \epsffile{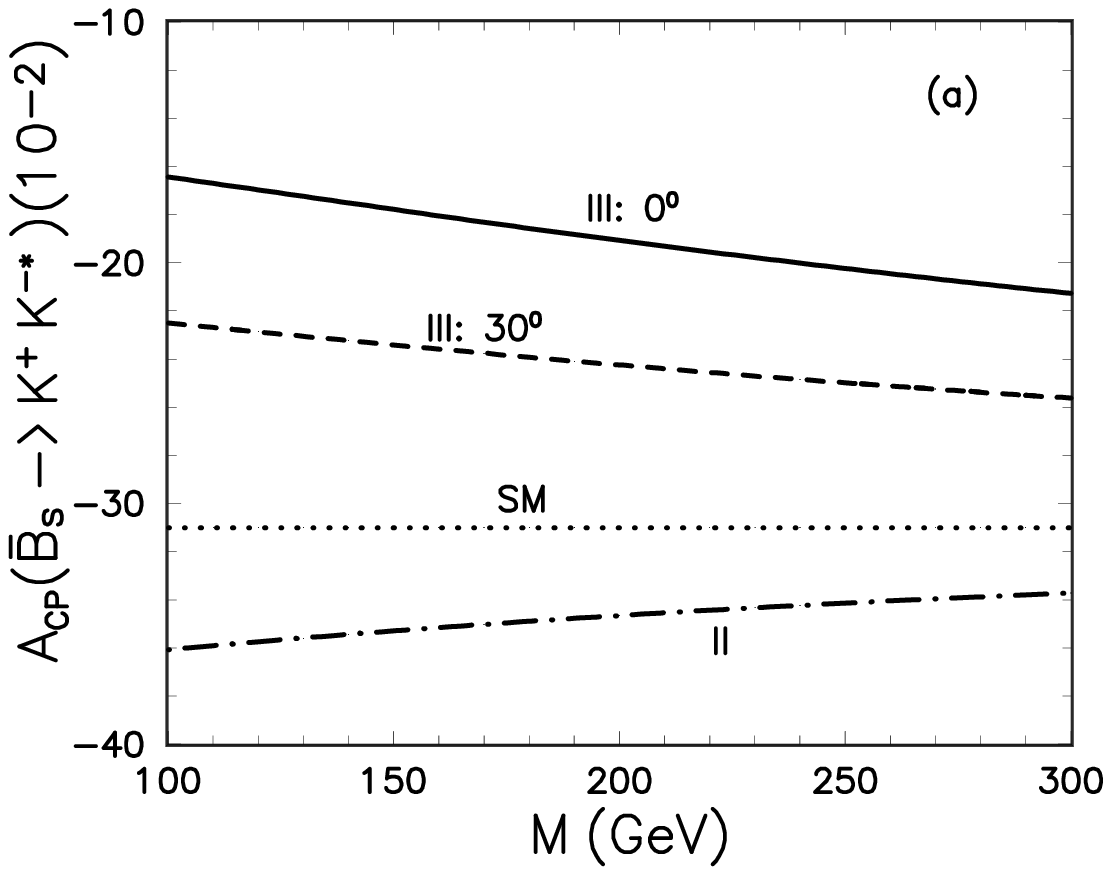}}
\vspace{-80pt}
\centerline{\epsfxsize=\textwidth \epsffile{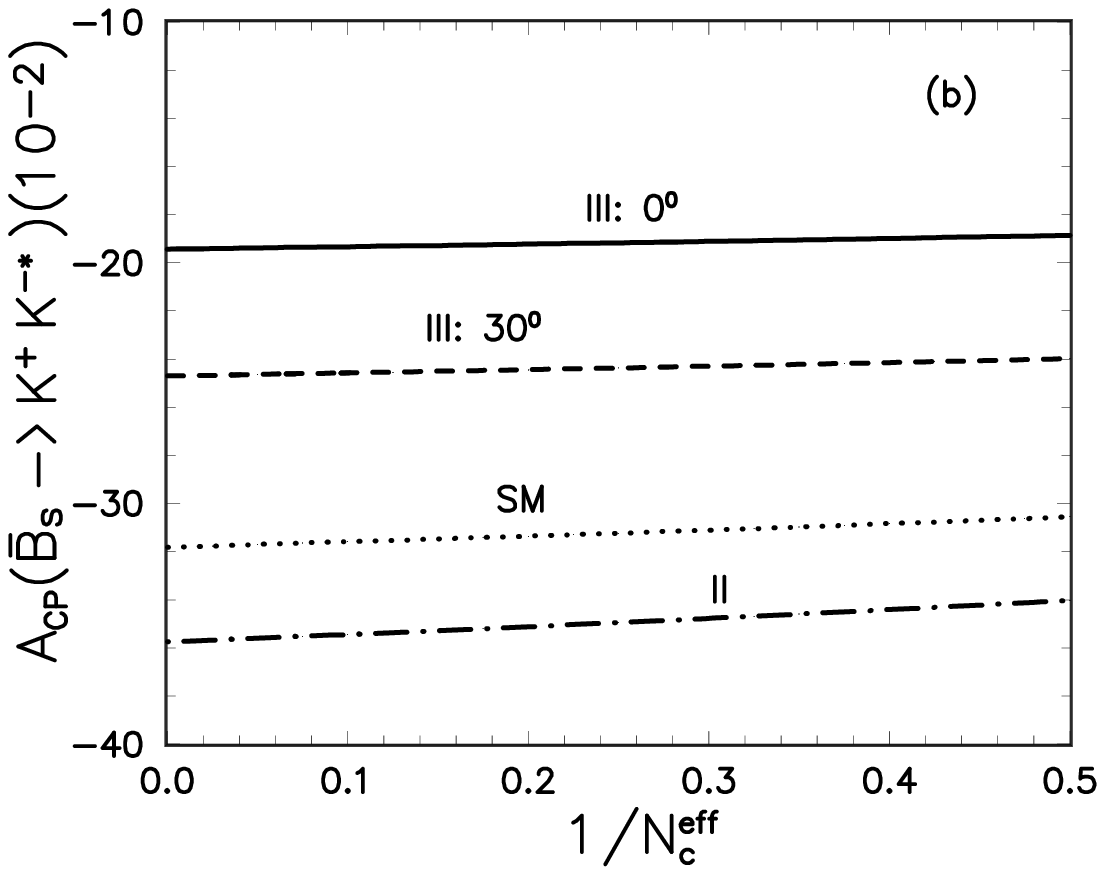}}
\vspace{-20pt}
\caption{Same as \fig{fig:fig6} but for decay $\overline{B}_s \to K^+ K^{-*}$.}
\label{fig:fig7}
\end{minipage}
\end{figure}

\begin{figure}[t] %fig.8
\vspace{-60pt}
\begin{minipage}[]{0.96\textwidth}
\centerline{\epsfxsize=\textwidth \epsffile{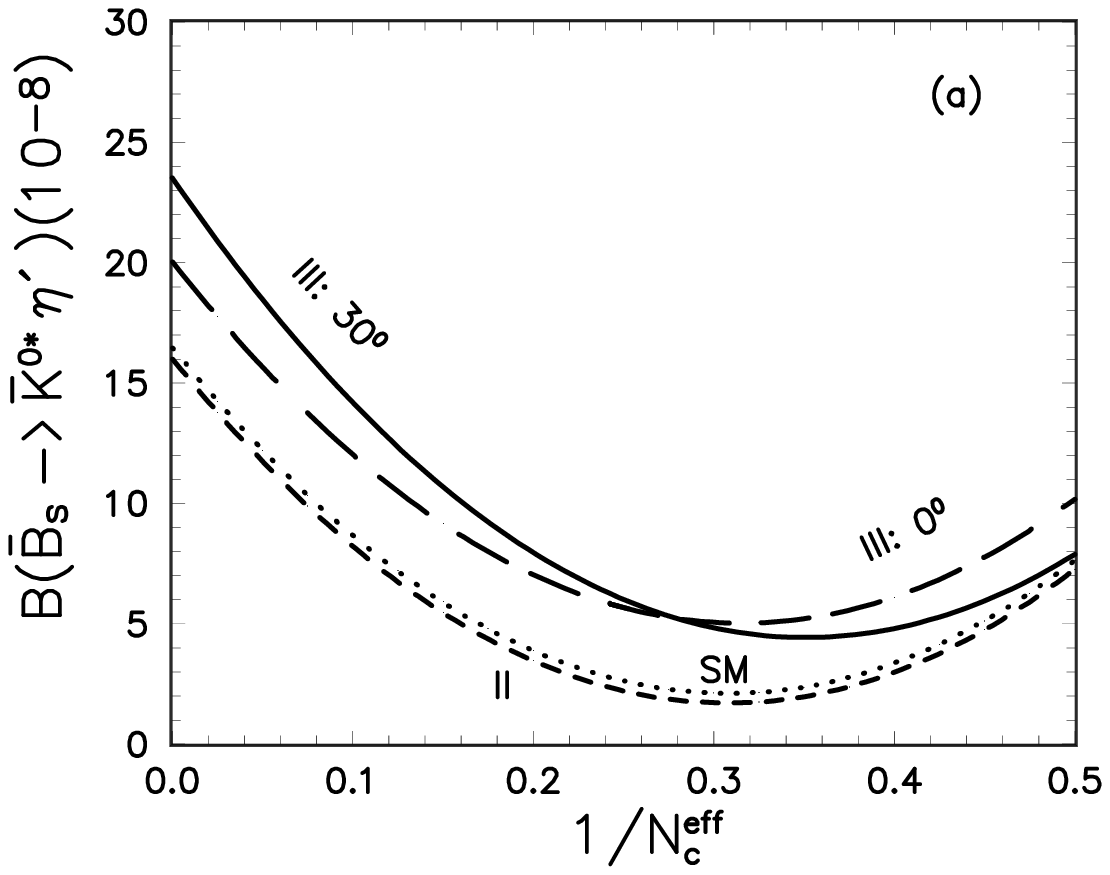}}
\vspace{-80pt}
\centerline{\epsfxsize=\textwidth \epsffile{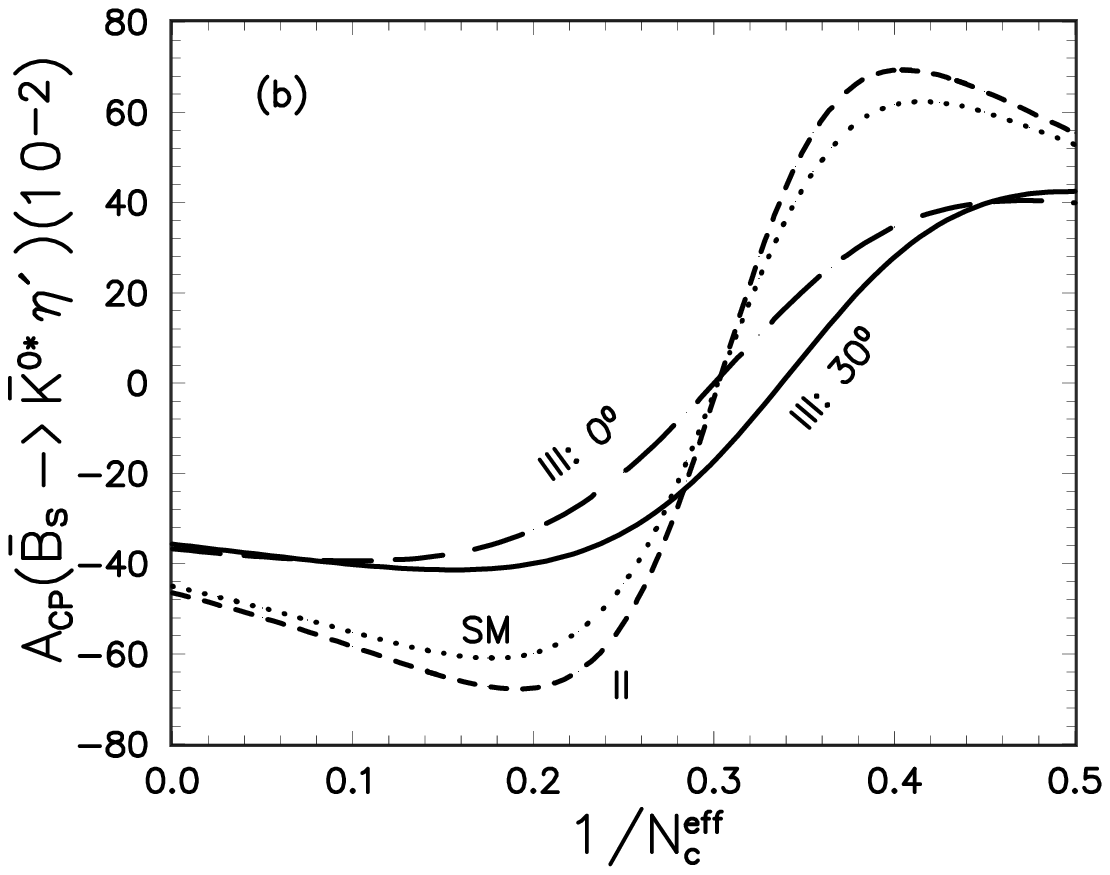}}
\vspace{-20pt}
\caption{ Branching ratios and CP-violating asymmetries of
 $\overline{B}_s \to K^{0*} \etap $ decay versus $1/\nceff$ in the SM and
 models II and III, assuming $\mhp=200$GeV and $\tan {\beta}=2$.
 The four curves correspond to the theoretical predictions in the SM
(dotted curve ), model II (short-dashed curve), model III with
$\theta=0^\circ$ (long-dashed curve) and $\theta=30^\circ$ (solid  curve),
respectively.}
\label{fig:fig8}
\end{minipage}
\end{figure}

In models I and II, the new physics corrections on $\acp$ for almost all
$B_s$ decay modes studied here are negligibly small  as can be seen from
Table \ref{acpm1} and Figures \ref{fig:fig6}-\ref{fig:fig8}.
In  model III, the new physics correction is varying from channel to
channel, as illustrated in Table \ref{acpsm1}  and Figures
\ref{fig:fig6}-\ref{fig:fig8}:
\begin{itemize}

\item
For $\ov B_s \to K^+ K^-$ decay, the new physics correction to its $\acp$ is
very small in size and insensitive to the variations of $\nceff$ and
$\theta$.

\item
For $\ov B_s \to K^+ K^{-*}$ decay, the new physics correction to its
$\acp$ is moderate in size, from $-20$ to $-40\%$ with $0^\circ \leq
\theta \leq 30^\circ$, and insensitive to the variations of $\nceff$.

\item
For $\ov B_s \to \bar{K}^{0*} \etap$ and three remaining decays given
in Eq.(\ref{eq:acp10}), the size and the sign of the new physics
corrections strongly depend on both $\nceff$ and $\theta$.

\item
For $\ov B_s \to \eta \etap, \phi \phi $ and several other CP-class-2 and 3
decays, the new physics corrections can be as large as a factor of
30, but have a very strong dependence on $\nceff$ and $\theta$. Despite
the large new physics correction to these decay modes, their $\acp$  are
still smaller than five percent because of strong suppression of $1/x^2$.

\end{itemize}

For the QCD-penguin dominated $\ov B_s \to K^+ K^{-*}$ decay, its decay
amplitude is proportional to the combination of large and $\nceff$ stable
coefficient $a_1$ and $a_4$ \cite{chenbs99},
\beq
{\cal M}(\ov B_s \to K^+ K^{-*} )\propto
 ~[ V_{ub}V^*_{ts}a_1 - V_{tb} V^*_{ts} (a_4 + a_{10}) ]~.\label{eq:kpkms}
\eeq
The imaginary part of ${\cal M}$ for $b \to s$ and $\bar{b} \to
\bar{s}$ transitions are very different, which in turn leads to a large
$A_{CP}$. Numerical result indeed shows that this decay  has a
large and $\nceff$ stable CP-violating asymmetry,
\beq
\acp(\obar B_s \to K^\pm K^{\mp*}) \approx -30\%
\eeq
for $2\leq \nceff \leq \infty$.
Another advantage of this decay mode is the large ($\sim 70\%$) new physics
enhancement to its branching ratio ${\cal B}(\ov B_s \to K^+ K^{-*})$
in model III, as illustrated in Fig.\ref{fig:fig7}. Taking into account
above facts,  this decay mode $\ov B_s \to K^+ K^{-*} $ seems
to be the ``best" channel to find CP
violation of $B_s$ system through studies of two-body charmless decays
of $B_s$ meson.

Since the tree-dominated $\ov B_s \to K^+ \pi^-$ decay mode has
a moderate CP-violating asymmetry ( $\sim 10\%$ ), a large branching
ratio ($\sim 7 \times 10^{-6}$ ), negligible new physics correction,
large detection efficiency \footnote{In general, the detect efficiency for the
two-body $B$ meson decays with charged final states is larger than that with
neutral final states by a factor 2 or 3.} and a very
weak $\nceff$ dependence, we therefore classify this decay mode as
one of the promising decay channels for discovering the CP violation
in $B_s$ system.

For the decay $\ov B_s \to \bar{K}^{0*} \etap$, although the SM prediction
of its $\acp$ can be large, but it is varying in the range of $-60\%$ to
$60\%$ due to the strong dependence on $\nceff$, as illustrated in
Fig.\ref{fig:fig8}. Another disadvantage for this decay is its small
branching ratio, $(0.02-0.16) \times 10^{-6}$, almost two orders smaller
than that of $\ov B_s \to K^+ \pi^-$ and $K^+ K^{-*}$ decays.

For the remaining five decay modes as given in Eq.(\ref{eq:acp10}),
although the size of
their $\acp$ can also be as large as $20$ to $30\%$, but these decays
can not be  "good" channels for discovering the CP violation in $B_s$
system because of the strong $\nceff$ dependence and very small branching
ratios.

In Figures \ref{fig:fig6}-\ref{fig:fig7}, we show the mass and $\nceff$
dependence of $\acp$ for $\ov B_s \to K^+ K^-$ and $K^+ K^{-*}$ decays.
In these figures, the dotted and dot-dashed curve refers to the theoretical
prediction in the SM and model II, while the solid and short-dashed curve
corresponds to the prediction in the model III for $\theta=0^\circ$ and
$30^\circ$ respectively. As can be seen from \fig{fig:fig7}, the
CP-violating asymmetry of $\ov B_s \to K^+ K^{-*}$ decay are large in size
and has weak or moderate dependence on $\mhp$, $\nce$ and $\theta$.

\section{Summary and discussions}\label{sec:sum}

In this paper, we calculated the branching ratios and CP-violating
asymmetries of two-body charmless hadronic decays of $B_s$ meson
in the standard model and the general two-Higgs-doublet models
(models I, II, and III) by employing the NLO effective Hamiltonian
with generalized factorization. In Sec.\ref{sec:heff}, we defined
the effective Wilson coefficients $C_i^{eff}$ with the inclusion of
new physics contributions, and presented the formulas needed to
calculate the branching ratios ${\cal B} (B_s \to h_1 h_2)$.

In Sec. \ref{sec:bh1h2}, we calculated the branching ratios for thirty nine
$B_s \to h_1 h_2$ decays in the SM and models I, II, and III, presented the
numerical results in Tables \ref{brbssm}-\ref{brbsm2} and displayed the
$\mhp$ and $\nceff$ dependence of several interesting decay modes in
Figures \ref{fig:fig1}-\ref{fig:fig5}.
From the numerical results, one can see the following

\begin{itemize}

\item
In models I and II, the new physics corrections to the decay rates of all
$B_s \to h_1 h_2$ decay modes  are small and will be masked by other larger
known theoretical uncertainties.

\item
In model III, the new physics corrections to QCD penguin-dominated decays
$B_s \to K^0\etapp, K^+ K^{-*}, \phi\phi, etc.$, are large in size,
from $~30\%$ to $~130\%$ $w.r.t$ the SM predictions, and
insensitive to the variations of the mass $\mhp$ and $\nceff$.
For tree- or  electroweak penguin-dominated decay modes as listed in
Eq.(\ref{eq:small}), however,  the new physics corrections are very
small in size: $\delta {\cal B} \leq 5\%$.

\item
For the decays $\ov B_s \to \eta \etap$ and $\ov B_s \to \etap \etap$,
analogue of $B_d \to K^0 \etap$ decay, the branching ratios are large but
in comparable size with other six decay modes listed
in Eq.(\ref{eq:large8}). The new physics enhancements to
${\cal B}(\ov B_s \to \eta \etap)$ and ${\cal B}(\ov B_s \to \etap \etap)$
are significant in size, $\sim 70\%$ in model III.

\item
For decay modes $B_s \to \pi^0 \phi, \phi \etap, \rho^0\phi, \omega \phi$
and $B_s \to \phi \phi$, the variation of the branching ratios induced by
using the BSW or LCSR form factors is about a factor of 2, but small or
moderate for all other decay modes. This feature remain basically unchanged
after inclusion of new physics contributions.

\item
For $B_s \to K^+ K^-$ and other decay modes as listed in
Eq.(\ref{eq:large8}), the branching ratios are at the level of
$(1-3)\times 10^{-5}$ in  the SM and model III. These decay modes will
be measurable at the future hadron colliders with large $b$ production.

\end{itemize}

In Sec. \ref{sec:acp}, we calculated the CP-violating asymmetries $\acp$
for thirty nine $B_s \to h_1 h_2$ decays in the SM and 2HDM's, presented the
numerical results in Tables \ref{acpsm1}-\ref{acpm1} and displayed the
$\mhp$ and $\nceff$ dependence of $\acp$ for several typical decay modes
in Figures \ref{fig:fig6}-\ref{fig:fig8}. From those
tables and figures, the following conclusions can be drawn:

\begin{itemize}

\item
For almost all $B_s$ decay modes,  the new physics corrections on $\acp$ are
negligibly small in models I and II. In  model III, the new physics
correction is varying from channel to channel,  and has a strong dependence
on the parameter $\nceff$  and the new phase $\theta$ for most decay modes.

\item
For twenty four CP-class-2 and 3 $B_s$ meson decay modes, their
CP-violating asymmetries are small, $|\acp| \leq 5\%$, due to the
strong $1/x^2$ suppression.

\item
Among the studied thirty nine $B_s$  meson decay modes, seven of them
can have a CP-violating asymmetry larger than $20\%$ in magnitude.

\item
The $\ov B_s \to K^+ K^{-*}$ decay has a large and $\nceff$- and
$\theta$-stable CP-violating asymmetry, $ \approx -30\%$, and a large
branching ratio.
This mode seems to be the ``best" channel to find CP violation of $B_s$
system through studies of two-body charmless decays of $B_s$ meson.
The tree-dominated $\ov B_s \to K^+ \pi^-$ decay is also
a promising decay channel for discovering the CP violation in $B_s$
system.

\end{itemize}

%\vspace{1cm}
\section*{ACKNOWLEDGMENTS}

Authors are very grateful to K.T. Chao, L.B.Guo, C.D. L\"u and Y.D. Yang for
helpful discussions. C.S. Li acknowledge the support by the National Natural
Science Foundation of China,  the State Commission of Science
and technology of China and the Doctoral Program Foundation of Institution
of Higher Education. Z.J. Xiao acknowledges the support by the National
Natural Science Foundation of China under Grant No.19575015 and
10075013, and the Excellent Young Teachers Program of Ministry of Education,
P.R.China.

%\newpage
\begin{appendix}
\section{Input parameters and form factors} \label{app:a}

In this appendix we present relevant input parameters. The input
parameters are similar with those used in Ref.\cite{chenbs99}.

\begin{itemize}

\item
The coupling constants, B meson masses, light
meson masses, $\cdots$,  are as follows (all masses in unit of GeV )
\cite{chenbs99,pdg2000}
\beq
\alpha_{em}&=&1/128., \;  \alpha_s(M_Z)=0.118,\;  \sin^2\theta_W=0.23,\;
G_F=1.16639\times 10^{-5} (GeV)^{-2}, \non
M_Z&=&91.188, \;   M_W=80.42,\;
m_{B_s^0}=5.369,\;   m_{\pi^\pm}=0.140,\;\non
m_{\pi^0}&=&0.135,\;   m_{\eta}=0.547,\; m_{\etap}=0.958,\;
m_{\rho}=0.770,\;  m_{\omega}=0.782,\non
m_{\phi}&=&1.019,\; m_{K^\pm}=0.494,\;  m_{K^0}=0.498,\;  m_{K^{*\pm}}=0.892,\non
m_{K^{*0}}&=&0.896, \; \tau(B_s^0)= 1.493 ps, \label{masses}
\eeq

\item
For the elements of CKM matrix, we use Wolfenstein parametrization and fix
the parameters $A, \lambda, \rho, \eta$ to their central values:
\beq
A=0.804,\; \lambda=0.22, \; \rho=0.16, \; \eta=0.34 \label{sip:ckm}
\eeq

\item
Following Refs.\cite{tseng99,chen99}, the current quark masses evaluated at
the scale $\mu= m_b$ will be used in the numerical calculations,
\beq
&&m_b(m_b)=4.34 GeV, \; m_c(m_b)=0.95 GeV, m_s(m_b)=0.105 GeV, \non
&& m_d(m_b)=6.4 MeV,\; m_u(m_b)=3.2 MeV. \label{cur-mass}
\eeq
For the mass of heavy top quark we also use $m_t=\overline{m_t}(m_t)=168 GeV$.

\item
For the decay constants of light mesons, the following values are used in
the numerical calculations (in the units of MeV):
\beq
&&f_{\pi}=133,\; f_{K}=160,\; f_{K^*}=221, \; f_{\rho}=210, \;
f_{\omega}=195, \; f_{\phi}=237, \; \non
&&f^u_{\eta}=f^d_{\eta}=78,\; f^u_{\etap}=f^d_{\etap}=63,\;
f^c_{\eta}=-2.4,\; f^c_{\etap}=-6.3,\non
&&f^s_{\eta}=-112,\; f^s_{\etap}=137,\; .
\label{fpis}
\eeq
where $f^u_{\etapp}$ and $f^s_{\etapp}$ have been defined in the
two-angle-mixing formalism with $\theta_0=-9.2^0$ and $\theta_8
=-21.2^0$\cite{fks98}.

\item
In the calculation we use the following BSW form factors $F(0)$(in the units of
GEV) \cite{du1993,bsw87,chenbs99},
\beq
&&F_0^{B\to \pi}(0)=0.33,\; F_0^{B\to K}(0)=0.274,\;
F_0^{B\to \eta}(0)=-0.212,\; F_0^{B\to \etap}(0)=0.218,\non
&&A_{0,1,2}^{B\to \phi}(0)=0.273,\;
A_{0}^{B\to K^*}(0)=0.236, \;A_{1,2}^{B\to K^*}(0)=0.232,\;\non
&& V^{B\to \phi}(0)=0.319, \; V^{B\to K^*}(0)=0.2817.
\label{eq:bsw-f0}
\eeq
We  use the monopole $k^2$-dependence for form factors,
\beq
f_i(k^2)&=& \frac{f_i(0)}{1-k^2/m^2_*},
\eeq
where $m_*$ is the pole mass given in \cite{bsw87}:
\beq
\{ m(0^-),m(1^-),m(1^+),m(0^+) \}&=& \{ 5.2789, 5.3248,5.37,5.73 \},
\eeq
for $\bar{u}b$ and $ \bar{d}b$ currents. And
\beq
\{ m(0^-),m(1^-),m(1^+),m(0^+)\}&=& \{5.3693, 5.41,5.82,5.89\},
\eeq
for $\bar{s}b $ currents.

\item
For the decays involving $B_s \to K^*$ and $B_s \to \phi$ transitions, we
also consider the case of using LCSR form factors with  the $k^2$-dependence
as defined in Ref.\cite{ball98},
\beq
f(k^2)=\frac{f(0)}{ 1- a (k^2/M_{B_s}^2) + b(k^2/M_{B_s}^2)^2},
\eeq
where the values of $f(0)$ and coefficients $a$ and $b$ have been given
in Ref.\cite{ball98}.

\end{itemize}
\end{appendix}

\newpage

\begin{table}%[tph]
\begin{center}
\caption{Numerical values of $a_i$ for the transitions $b \to d$ [$\bar{b}
\to \bar{d}$ ].  The first, second and third entries for $a_3, \ldots, a_{10}$
refer to the values of $a_i$ in the SM and models II and III, respectively.
All entries for $a_3, \ldots, a_{10}$ should be multiplied with $10^{-4}$. }
\label{ai:bd}
\vspace{0.2cm}
\begin{tabular} {lccc}
        &$\nceff=2$               & $\nceff=3$                   & $\nceff=\infty$
        \\ \hline
 $a_1$  &$0.985 \;[0.985]$            &$1.046\;[1.046]$           & $1.169  \;[1.169]$        \\
 $a_2$  &$0.216 \;[0.216]$            &$0.021\;[0.021]$           & $-0.369 \;[-0.369]$       \\
 $a_3$  &$-10.4-19.1i\;[-11.5-25.7i]$ &$66.1  \;[66.1]$           & $219+38.1i\;[221+51.4i]$      \\
        &$-33.0-19.1i\;[-34.1-25.7i]$ &$66.2  \;[66.2]$           & $265+38.1i\;[267+51.4i]$      \\
 $a_4$  &$-349-95.3i\;[-354-129i]$    &$-386-102i\;[-392-137i]$   & $-459-114i\;[-466-154i]$   \\
        &$-463-95.3i\;[-469-129i]$    &$-507-102i\;[-513-137i]$   & $-596-114i\;[-602-154i]$   \\
 $a_5$  &$-163-19.1i \;[-164-25.7i]$  &$-61.5    \;[-61.5]$       & $142+ 38.1i\;[144 + 51.4i]$   \\
        &$-186-19.1i \;[-187-25.7i]$  &$-61.4    \;[-61.4]$       & $187+ 38.1i\;[189 + 51i.4]$   \\
$a_6$   &$-538-95.3i\;[-544-129i]$    &$-562-102i\;[-568-137i]$   & $-609-114i\;[-616-154i]$   \\
        &$-652-95.3i\;[-657-129i]$    &$-683-102i\;[-689-137i]$   & $-746-114i\;[-752-154i]$   \\
 $a_7$  &$5.2-2.5i\;[5.1-3.1i]$       &$4.1-2.5i\;[4.0-3.1i]$     & $2.1-2.5i\;[2.0-3.1i]$    \\
        &$5.4-2.5i\;[5.3-3.1i]$       &$4.3-2.5i\;[4.2-3.1i]$     & $2.2-2.5i\;[2.1-3.1i]$    \\
 $a_8$  &$7.2-1.3i\;[7.2-1.6i]$       &$6.9-0.8i \;[6.8-1.0i]$    & $6.2     \;[6.2]$         \\
        &$7.4-1.3i\;[7.3-1.6i]$       &$7.0-0.8i \;[7.0-1.0i]$    & $6.3    \;[6.3]$          \\
 $a_9$  &$-85.8-2.5i\;[-85.9-3.1i]$   &$-91.7-2.5i\;[-91.8-3.1i]$ & $-103-2.5i\;[-104-3.1i]$\\
        &$-86.4-2.5i\;[-86.5-3.1i]$   &$-92.3-2.5i\;[-92.4-3.1i]$ & $-104.1-2.5i\;[-104-3.1i]$      \\
$a_{10}$&$-16.5-1.3i\;[-16.6-1.6i]$   &$0.7 -0.8i\;[0.7-1.0i]$    & $35.2     \;[35.2]$           \\
        &$-16.6-1.3i\;[-16.7-1.6i]$   &$0.7 -0.8i\;[0.7-1.0i]$    & $35.4     \;[35.4]$         \\
\end{tabular}\end{center}
\end{table}

\begin{table}%[tph]
\begin{center}
\caption{Same as \tab{ai:bd} but for $b \to s$ [$\bar{b} \to \bar{s}$ ]
transitions. }
\label{ai:bs}
\vspace{0.2cm}
\begin{tabular} {lccc}
       & $\nceff=2$                    & $\nceff=3$                    & $\nceff=\infty$  \\ \hline
 $a_1$ &$0.985  \;[0.985]$          &$1.046\;[1.046]$            &$1.169  \;[1.169]$       \\
 $a_2$ &$0.216  \;[0.216]$          &$0.021\;[0.021]$            &$-0.369 \;[-0.369]$ \\
 $a_3$ &$-10.9-21.7i\;[-9.8-22.1i]$ &$66.1   \;[66.1]$          &$220+43.3i\;[218+44.3i]$   \\
       &$-33.6-22.7i\;[-32.5-22.2i]$&$66.2  \;[66.2]$           &$266+43.3i\;[264+44.3i]$       \\
 $a_4$ &$-352-108i\;[-346-111i]$  &$-389-116i\;[-383-118i]$    &$-462-130i\;[-455-133i]$\\
       &$-467-108i\;[-460-111i]$  &$-510-116i\;[-504-118i]$    &$-599-130i\;[-592-133i]$       \\
 $a_5$ &$-164-22.7i\;[-162-22.2i]$&$-61.5     \;[-61.5]$       &$143+ 43.3i\;[140 + 44.3i]$  \\
       &$-186-21.7i\;[-185-22.2i]$&$-61.4     \;[-61.4]$       &$188+ 43.3i\;[186 + 44.3i]$\\
 $a_6$ &$-541-108i\;[-535-111i]$  &$-565-116i\;[-559-118i]$    &$-612-130i\;[-606-133i]$  \\
       &$-654-108i\;[-649-111i]$  &$-686-116i\;[-680-118i]$    &$-749-130i\;[-742-133i]$ \\
 $a_7$ &$5.1-2.8i \;[5.2-2.8i]$   &$4.1-2.8i\;[4.2-2.8i]$      &$2.0-2.8i\;[2.1-2.8i]$   \\
       &$5.3-2.8i \;[5.4-2.8i]$   &$4.3-2.8i\;[4.4-2.8i]$      &$2.2-2.8i\;[2.3-2.8i]$  \\
 $a_8$ &$7.2-1.4i \;[7.2-1.4i]$   &$6.9-0.9i  \;[6.9-0.9i]$    &$6.2      \;[6.2]$    \\
       &$7.4-1.4i \;[7.4-1.4i]$   &$7.0-0.9i  \;[7.0-0.9i]$    &$6.3      \;[6.3]$   \\
 $a_9$ &$-85.9-2.8i\;[-85.8-2.8i]$&$-91.7-2.8i\;[-91.6-2.8i]$  &$-104-2.8i\;[-103-2.8i]$     \\
       &$-86.5-2.8i \;[-86.4-2.8i]$   &$-92.4-2.8i\;[-92.3-2.8i]$ &$-104-2.8i\;[-104-2.8i]$    \\
$a_{10}$&$-16.6-1.4i\;[-16.5-1.4i]$   &$0.7-0.9i \;[0.7-0.9i]$ &$35.2\;[35.2]$                \\
        &$-16.7-1.4i\;[-16.6-1.4i]$   &$0.7-0.9i \;[0.7-0.9i]$ &$35.4\;[35.4]$                 \\
\end{tabular}\end{center}
\end{table}

\newpage
\begin{table}[tb]
\begin{center}
\caption{${\cal B}(B_s \to h_1, h_2)$ (in units of $10^{-6}$) in the SM and model
III by using the BSW form factors, and assuming $k^2=m_b^2/2$, $\rho=0.16$, $\eta=0.34$,
$\mhp=200$GeV, $\theta=0^\circ$, and $\nceff=2,\; 3,\; \infty$. }
\label{brbssm}
\vspace{0.2cm}
\begin{tabular}{ll rrr rrr rrr}
& & \multicolumn{3}{c}{SM: $\ \ \ {\cal B}$ }
& \multicolumn{6}{c}{Model III: $\ \ \ {\cal B}$ and $\delta {\cal B}[\%]$ }
\\
Channel                  & Class &$2$ & $3$& $\infty$&$2$ & $3$& $\infty$ &$2$ & $3$& $\infty$   \\ \hline
$\bsb \to K^+ \pi^-$     & I & $6.33$ &$7.14$&$8.89$& $6.52$ &$7.35$&$9.16$& $3.1 $ &$3.1$ &$3.0$ \\
$\bsb \to K^0 \pi^0$     & II& $0.19$ &$0.08$&$0.56$& $0.24$ &$0.14$&$0.64$& $23.8$ &$67.1$&$14.4$  \\
$\bsb \to K^0 \eta $     & VI& $0.34$ &$0.31$&$0.79$& $0.47$ &$0.46$&$1.00$& $38.3$ &$49.9$&$26.8$ \\
$\bsb \to K^0 \etap $    & VI& $0.57$ &$0.51$&$0.77$& $0.88$ &$0.84$&$1.17$& $53.0$ &$65.6$&$52.3$ \\
$\bsb \to K^+   K^-$     & IV& $10.7$ &$11.7$&$14.0$& $16.7$ &$18.5$&$22.3$& $56.5$ &$57.6$&$59.4$ \\
$\bsb \to \pi^0 \eta$    & V & $0.04$ &$0.06$&$0.11$& $0.04$ &$0.06$&$0.11$& $1.9 $ &$1.8 $ &$1.3$ \\
$\bsb \to \pi^0 \etap$   & V & $0.04$ &$0.06$&$0.10$& $0.04$ &$0.06$&$0.11$& $1.9 $ &$1.8 $ &$1.3$ \\
$\bsb \to \eta \etap$    & VI& $13.8$ &$15.9$&$20.5$& $22.5$ &$25.9$&$33.4$& $63.8$ &$63.3$&$62.6$ \\
$\bsb \to \etap\etap$    & VI& $6.79$ &$7.51$&$9.08$& $11.6$ &$12.9$&$15.7$& $70.6$ &$71.7$&$73.4$ \\
$\bsb \to \eta \eta$     & VI& $6.97$ &$8.37$&$11.6$& $10.9$ &$13.0$&$17.7$& $56.9$ &$55.3$&$52.8$ \\
$\bsb \to K^0 \bar{K}^0$ & IV& $11.4$ &$13.2$&$17.3$& $17.6$ &$20.4$&$26.4$& $66.2$ &$65.6$&$64.5$ \\
\hline
$\bsb \to K^{*+} \pi^-$       &I  &$4.04$ &$4.56$&$5.70 $&$4.04$ &$4.56$&$5.70 $&$ 0.0$ &$ 0.0$&$ 0.0$ \\
$\bsb \to K^{+} \rho^-$      &I   &$14.8$ &$16.7$&$20.8 $&$14.9$ &$16.8$&$21.0 $&$ 0.9$ &$ 0.9$&$ 0.9$ \\
$\bsb \to K^{0*} \pi^0$       &II &$0.10$ &$0.003$&$0.29$&$0.10$ &$0.002$&$0.29$&$-1.7$ &$-36.3$&$ 0.1$ \\
$\bsb \to K^{0} \rho^0$      &II  &$0.35$ &$0.04$&$1.11 $&$0.37$ &$0.07$&$1.17 $&$ 6.8$ &$ 93.8$&$ 5.3$ \\
$\bsb \to K^{0} \omega$    &II,VI &$1.14$ &$0.16$&$1.81 $&$1.42$ &$0.26$&$1.83 $&$ 24.7$ &$ 56.7$&$ 1.2$ \\
$\bsb \to \bar{K}^{0*} \eta $    &II,VI &$0.16$ &$0.13$&$0.44 $&$0.22$ &$0.21$&$0.55 $&$ 38.4$ &$ 58.2$&$ 24.9$ \\
$\bsb \to \bar{K}^{0*} \etap$    &II,VI &$0.08$ &$0.02$&$0.16 $&$0.10$ &$0.05$&$0.20 $&$ 33.6$ &$ 131$&$ 21.7$ \\
$\bsb \to K^{+} K^{-*}$       &IV &$3.05$ &$3.39$&$4.12 $&$5.03$ &$5.61$&$6.86 $&$ 64.7$ &$ 65.3$&$ 66.4$ \\
$\bsb \to K^{+*}K^{-} $       &IV &$0.89$ &$0.97$&$1.15 $&$0.90$ &$0.99$&$1.18 $&$ 2.2$ &$ 2.3$&$ 2.5$ \\
$\bsb \to \rho \eta   $       &V  &$0.08$ &$0.11$&$0.25 $&$0.08$ &$0.12$&$0.25 $&$ 1.0$ &$ 1.0$&$ 0.8$ \\
$\bsb \to \rho \etap  $       &V  &$0.08$ &$0.11$&$0.24 $&$0.08$ &$0.11$&$0.24 $&$ 1.0$ &$ 1.0$&$ 0.8$ \\
$\bsb \to \omega \eta $       &V  &$0.85$ &$0.01$&$2.60 $&$1.29$ &$0.01$&$4.15 $&$ 51.5$ &$-1.4$&$ 59.9$ \\
$\bsb \to \omega \etap$       &V  &$0.84$ &$0.01$&$2.56 $&$1.28$ &$0.01$&$4.09 $&$ 51.5$ &$-1.4$&$ 59.9$ \\
$\bsb \to \pi^0 \phi  $       &V  &$0.13$ &$0.17$&$0.32 $&$0.13$ &$0.17$&$0.32 $&$ 1.9$ &$ 1.8$&$ 1.3$ \\
$\bsb \to \phi \eta   $       &VI &$1.85$ &$0.76$&$0.07 $&$3.78$ &$1.69$&$0.03 $&$ 104$ &$ 122$&$-53.5$ \\
$\bsb \to \phi \etap  $       &VI &$0.70$ &$0.20$&$1.49 $&$1.82$ &$0.40$&$1.14 $&$ 161$ &$ 107$&$-23.5$ \\
$\bsb \to K^0 \bar{K}^{0*}$   &IV &$3.24$ &$4.11$&$6.17 $&$5.52$ &$6.85$&$9.93 $&$ 70.6$ &$ 66.7$&$ 61.0$ \\
$\bsb \to K^{0*}\bar{K}^0$    &IV &$0.39$ &$0.31$&$0.18 $&$0.40$ &$0.32$&$0.19 $&$ 0.8$ &$ 0.9$&$ 1.0$ \\
$\bsb \to K^0 \phi    $       &VI &$0.001$&$0.03$&$0.30 $&$0.004$&$0.03$&$0.40 $&$ 118$ &$ 1.1$&$ 38.4$ \\
\hline
$\bsb \to K^{+*} \rho^-$       &I  &$12.5$ &$14.1$&$17.5$&$12.6$ &$14.2$&$17.7 $&$0.9 $ &$ 0.9 $&$0.9 $ \\
$\bsb \to K^{0*} \rho^0$       &II &$0.29$ &$0.03$&$0.94$&$0.31$ &$0.06$&$0.99 $&$6.8 $ &$ 93.8$&$5.3 $ \\
$\bsb \to K^{0*} \omega$     &II,VI&$0.24$ &$0.03$&$0.38$&$0.30$ &$0.05$&$0.39 $&$24.7$ &$ 56.7$&$1.2 $ \\
$\bsb \to K^{+*} K^{-*}$       &IV &$2.72$ &$3.02$&$3.68$&$4.48$ &$5.00$&$6.12 $&$64.7$ &$ 65.3$&$66.4$ \\
$\bsb \to \rho^0 \phi  $       &V  &$0.15$ &$0.21$&$0.45$&$0.15$ &$0.21$&$0.46 $&$1.0 $ &$ 0.99$&$0.8 $ \\
$\bsb \to \omega \phi  $       &V  &$0.79$ &$0.01$&$2.41$&$1.20$ &$0.01$&$3.85 $&$51.3$ &$-1.35$&$59.9$ \\
$\bsb \to K^{0*}\bar{K}^{0*}$  &IV &$2.14$ &$2.71$&$4.07$&$3.65$ &$4.53$&$6.56 $&$70.7$ &$ 66.8$&$61.1$ \\
$\bsb \to K^{0*}\phi   $       &VI &$0.03$ &$0.12$&$0.48$&$0.05$ &$0.19$&$0.74 $&$68.5$ &$ 58.9$&$54.1$ \\
$\bsb \to \phi  \phi   $       &VI &$17.5$ &$8.99$&$0.42$&$29.9$ &$15.8$&$0.98 $&$71.1$ &$ 75.8$&$134 $ \\
\end{tabular}\end{center}
\end{table}

\newpage
\begin{table}[tb]
\begin{center}
\caption{${\cal B}(B_s \to PV, VV)$ (in units of $10^{-6}$) in the SM and model
III by using the LCSR form factors for $B_s \to K^*$ or $B_s \to \phi$ transition,
and assuming $k^2=m_b^2/2$, $\rho=0.16$, $\eta=0.34$,
$\mhp=200$GeV, $\theta=0^\circ$, and $\nceff=2,\; 3,\; \infty$. }
\label{brbslcr}
\vspace{0.2cm}
\begin{tabular}{ll rrr rrr rrr}
& & \multicolumn{3}{c}{SM: $\ \ \ {\cal B}$ }
& \multicolumn{6}{c}{Model III: $\ \ \ {\cal B}$ and $\delta {\cal B}[\%]$ }
\\
Channel                  & Class &$2$ & $3$& $\infty$&$2$ & $3$& $\infty$ &$2$ & $3$& $\infty$   \\ \hline
$\bsb \to K^{*+} \pi^-$       &I  &$4.68$ &$5.29$ &$6.61$ &$4.69$&$5.29$ &$6.61$&$ 0.0$ &$ 0.0 $&$ 0.0$ \\
$\bsb \to K^{+} \rho^-$       &I  &$15.4$ &$17.4$ &$21.7$ &$15.5$&$17.5$ &$21.9$&$ 0.9$ &$ 0.9 $&$ 0.9$ \\
$\bsb \to K^{0*} \pi^0$       &II &$0.12$ &$0.003$&$0.33$ &$0.11$&$0.002$&$0.33$&$-1.7$ &$-36.3$&$ 0.1$ \\
$\bsb \to K^{0} \rho^0$      &II  &$0.36$ &$0.04$ &$1.16$ &$0.39$&$0.07$ &$1.22$&$ 6.8$ &$ 93.8$&$ 5.3$ \\
$\bsb \to K^{0} \omega$    &II,VI &$1.19$ &$0.17$ &$1.89$ &$1.49$&$0.27$ &$1.91$&$ 24.7$&$ 56.7$&$ 1.2$ \\
$\bsb \to \bar{K}^{0*} \eta $    &II,VI &$0.18$ &$0.14$ &$0.50$ &$0.24$&$0.22$ &$0.62$&$ 36.7$&$ 57.8$&$ 23.6$ \\
$\bsb \to \bar{K}^{0*} \etap$    &II,VI &$0.09$ &$0.02$ &$0.19$ &$0.11$&$0.05$ &$0.23$&$ 28.7$&$ 137$&$ 17.5$ \\
$\bsb \to K^{+} K^{-*}$       &IV &$3.22$ &$3.58$ &$4.35$ &$5.31$&$5.92$ &$7.22$&$ 64.7$&$ 65.3$&$ 66.4$ \\
$\bsb \to K^{+*}K^{-} $       &IV &$1.04$ &$1.14$ &$1.35$ &$1.06$&$1.16$ &$1.38$&$ 2.2$ &$ 2.3$&$ 2.8$ \\
$\bsb \to \rho \eta   $       &V  &$0.09$ &$0.12$ &$0.26$ &$0.09$&$0.12$ &$0.26$&$ 1.0$ &$ 1.0$&$ 0.8$ \\
$\bsb \to \rho \etap  $       &V  &$0.09$ &$0.12$ &$0.25$ &$0.09$&$0.12$ &$0.25$&$ 1.0$ &$ 1.0$&$ 0.8$ \\
$\bsb \to \omega \eta $       &V  &$0.89$ &$0.01$ &$2.71$ &$1.35$&$0.01$ &$4.33$&$ 51.5$&$-1.4$&$ 59.9$ \\
$\bsb \to \omega \etap$       &V  &$0.88$ &$0.01$ &$2.67$ &$1.33$&$0.01$ &$4.27$&$ 51.5$&$-1.4$&$ 59.9$ \\
$\bsb \to \pi^0 \phi  $       &V  &$0.26$ &$0.33$ &$0.63$ &$0.26$&$0.34$ &$0.64$&$ 1.9$ &$ 1.8$&$ 1.3$ \\
$\bsb \to \phi \eta   $       &VI &$1.36$ &$0.49$ &$0.18$ &$3.04$&$1.23$ &$0.09$&$ 124$ &$ 151$&$-50.3$ \\
$\bsb \to \phi \etap  $       &VI &$0.38$ &$0.53$ &$3.43$ &$0.87$&$0.21$ &$2.91$&$ 127$ &$-60.3$&$-15.3$ \\
$\bsb \to K^0 \bar{K}^{0*}$   &IV &$3.42$ &$4.34$ &$6.52$ &$5.83$&$7.23$ &$10.5$&$ 70.6$ &$66.7$&$ 61.0$ \\
$\bsb \to K^{0*}\bar{K}^0$    &IV &$0.46$ &$0.37$ &$0.22$ &$0.46$&$0.37$ &$0.22$&$ 0.8$ &$ 0.9 $&$ 1.0$ \\
$\bsb \to K^0 \phi    $       &VI &$0.004$&$0.05$ &$0.36$ &$0.002$&$0.05$&$0.50$&$-56.3$ &$ 1.1$&$ 36.0$ \\
$\bsb \to K^{+*} \rho^-$      &I  &$13.2$ &$14.9$&$18.6$ &$13.3$ &$15.0$&$18.8$ &$0.9$ &$ 0.9  $&$0.9$ \\
$\bsb \to K^{0*} \rho^0$      &II &$0.31$ &$0.03$&$0.99$ &$0.33$ &$0.06$&$1.05$ &$6.8$ &$ 93.8 $&$5.3$ \\
$\bsb \to K^{0*} \omega$    &II,VI&$0.26$ &$0.04$&$0.40$ &$0.32$ &$0.06$&$0.41$ &$24.7$ &$ 56.7$&$1.2$ \\
$\bsb \to K^{+*} K^{-*}$      &IV &$2.82$ &$3.13$&$3.79$ &$4.64$ &$5.17$&$6.33$ &$64.7$ &$ 65.3$&$66.4$ \\
$\bsb \to \rho^0 \phi  $      &V  &$0.27$ &$0.38$&$0.82$ &$0.28$ &$0.38$&$0.82$ &$1.0$ &$ 1.0  $&$0.8$ \\
$\bsb \to \omega \phi  $      &V  &$1.43$ &$0.01$&$4.33$ &$2.16$ &$0.01$&$6.93$ &$51.5$ &$-1.4 $&$59.9$ \\
$\bsb \to K^{0*}\bar{K}^{0*}$ &IV &$2.20$ &$2.80$&$4.20$ &$3.76$ &$4.67$&$6.77$ &$70.7$ &$ 66.8$&$61.1$ \\
$\bsb \to K^{0*}\phi   $      &VI &$0.07$ &$0.20$&$0.66$ &$0.12$ &$0.32$&$1.03$ &$68.9$ &$ 60.5$&$55.1$ \\
$\bsb \to \phi  \phi   $      &VI &$29.9$ &$15.4$&$0.72$ &$51.1$ &$27.0$&$1.68$ &$71.1$ &$ 75.8$&$134$ \\
\end{tabular}\end{center}
\end{table}

\begin{table}[tb]
\begin{center}
\caption{ ${\cal B}(B_s\to h_1 h_2)$ (in units of $10^{-6}$) in model I,
with $k^2=m_b^2/2$, $\rho=0.16$, $\eta=0.34$,
$\mhp=200$GeV, $\tan{\beta}=2$, and $\nceff=2,\; 3,\; \infty$. }
\label{brbsm1}
\vspace{0.2cm}
\begin{tabular}{ll rrr rrr rrr}
& & \multicolumn{3}{c}{SM: $\ \ \ {\cal B}$ }
& \multicolumn{6}{c}{Model I: $\ \ \ {\cal B}$ and $\delta {\cal B}[\%]$ } \\
Channel                       & Class &$2$ & $3$& $\infty$&$2$ & $3$& $\infty$ &$2$ & $3$& $\infty$   \\ \hline
$\bsb \to Ki^+ \pi^-$         & I &$6.33$ &$7.13$&$8.88 $&$6.34$ &$7.14$&$8.90$&$ 0.1$ &$ 0.1$&$0.1$ \\
$\bsb \to K^0 \pi^0$          & II&$0.19$ &$0.08$&$0.56 $&$0.19$ &$0.08$&$0.56$&$-0.1$ &$ 0.0$&$0.1$  \\
$\bsb \to K^0 \eta $          & VI&$0.34$ &$0.31$&$0.78 $&$0.34$ &$0.31$&$0.79$&$ 0.4$ &$ 0.7$&$0.4$ \\
$\bsb \to K^0 \etap $         & VI&$0.57$ &$0.51$&$0.76 $&$0.58$ &$0.52$&$0.77$&$ 1.2$ &$ 1.5$&$1.3$ \\
$\bsb \to K^+  K^-$           & IV&$10.6$ &$11.7$&$14.0 $&$10.8$ &$11.9$&$14.1$&$ 1.4$ &$ 1.3$&$1.3$ \\
$\bsb \to \pi^0 \eta$         & V &$0.04$ &$0.06$&$0.11 $&$0.05$ &$0.06$&$0.11$&$10.5$ &$10.0$&$7.3$ \\
$\bsb \to \pi^0 \etap$        & V &$0.04$ &$0.05$&$0.10 $&$0.05$ &$0.06$&$0.11$&$10.5$ &$10.0$&$7.3$ \\
$\bsb \to \eta \etap$         & VI&$13.7$ &$15.8$&$20.5 $&$13.9$ &$16.1$&$20.8$&$ 1.1$ &$ 1.2$&$1.3$ \\
$\bsb \to \etap\etap$         & VI&$6.77$ &$7.48$&$9.05 $&$6.89$ &$7.63$&$9.22$&$ 1.5$ &$ 1.5$&$1.6$ \\
$\bsb \to \eta \eta$          & VI&$6.95$ &$8.35$&$11.5 $&$7.03$ &$8.44$&$11.7$&$ 0.8$ &$ 0.9$&$1.0$ \\
$\bsb \to K^0 \bar{K}^0$      & IV&$11.4$ &$13.2$&$17.2 $&$11.6$ &$13.4$&$17.5$&$ 1.8$ &$ 1.9$&$2.0$ \\
\hline
$\bsb \to K^{*+} \pi^-$       &I  &$4.04$ &$4.56$&$5.70 $&$4.04$ &$4.56$&$5.70$&$ 0.0$ &$ 0.0$&$ 0.0 $ \\
$\bsb \to K^{+} \rho^-$       &I  &$14.7$ &$16.6$&$20.8 $&$14.8$ &$16.7$&$20.8$&$ 0.0$ &$ 0.0$&$ 0.0 $ \\
$\bsb \to K^{0*} \pi^0$       &II &$0.10$ &$0.003$&$0.29$&$0.10$&$0.003$&$0.29$&$ 0.3$ &$ 5.1$&$ 0.0 $ \\
$\bsb \to K^{0} \rho^0$      &II  &$0.35$ &$0.04$&$1.11 $&$0.35$ &$0.04$&$1.11$&$ 0.0$ &$ 0.5$&$ 0.0 $ \\
$\bsb \to K^{0} \omega$    &II,VI &$1.14$ &$0.16$&$1.81 $&$1.15$ &$0.17$&$1.81$&$ 0.6$ &$ 1.3$&$ 0.0 $ \\
$\bsb \to \bar{K}^{0*} \eta $    &II,VI &$0.16$ &$0.13$&$0.44 $&$0.16$ &$0.13$&$0.45$&$ 0.4$ &$ 0.8$&$ 0.5 $ \\
$\bsb \to \bar{K}^{0*} \etap$    &II,VI &$0.08$ &$0.02$&$0.16 $&$0.08$ &$0.02$&$0.17$&$ 0.6$ &$ 2.8$&$ 0.5 $ \\
$\bsb \to K^{+} K^{-*}$       &IV &$3.04$ &$3.38$&$4.11 $&$3.12$ &$3.45$&$4.18$&$ 2.1$ &$ 1.9$&$ 1.5 $ \\
$\bsb \to K^{+*}K^{-} $       &IV &$0.89$ &$0.97$&$1.15 $&$0.87$ &$0.96$&$1.14$&$-1.5$ &$-1.0$&$-0.2 $ \\
$\bsb \to \rho \eta   $       &V  &$0.08$ &$0.11$&$0.25 $&$0.09$ &$0.12$&$0.26$&$ 5.4$ &$ 5.6$&$ 4.6 $ \\
$\bsb \to \rho \etap  $       &V  &$0.08$ &$0.11$&$0.24 $&$0.09$ &$0.12$&$0.25$&$ 5.4$ &$ 5.6$&$ 4.6 $ \\
$\bsb \to \omega \eta $       &V  &$0.85$ &$0.01$&$2.59 $&$0.86$ &$0.01$&$2.65$&$ 0.8$ &$-7.4$&$ 2.0 $ \\
$\bsb \to \omega \etap$       &V  &$0.84$ &$0.01$&$2.55 $&$0.85$ &$0.01$&$2.61$&$ 0.8$ &$-7.4$&$ 2.0 $ \\
$\bsb \to \pi^0 \phi  $       &V  &$0.13$ &$0.17$&$0.32 $&$0.14$ &$0.18$&$0.34$&$ 10.5$&$ 10.0$&$ 7.3 $ \\
$\bsb \to \phi \eta   $       &VI &$1.84$ &$0.75$&$0.07 $&$1.86$ &$0.76$&$0.07$&$ 0.4$ &$ 0.2$&$ 2.6 $ \\
$\bsb \to \phi \etap  $       &VI &$0.69$ &$0.20$&$1.49 $&$0.71$ &$0.20$&$1.51$&$ 1.0$ &$ 0.0$&$ 1.2 $ \\
$\bsb \to K^0 \bar{K}^{0*}$   &IV &$3.22$ &$4.09$&$6.15 $&$3.30$ &$4.19$&$6.31$&$ 2.0$ &$ 2.1$&$ 2.3 $ \\
$\bsb \to K^{0*}\bar{K}^0$    &IV &$0.39$ &$0.31$&$0.18 $&$0.39$ &$0.31$&$0.18$&$-0.4$ &$-1.0$&$-2.9 $ \\
$\bsb \to K^0 \phi    $       &VI &$0.001$&$0.03$&$0.30 $&$0.002$&$0.03$&$0.30$&$ 0.3$ &$ 2.1$&$ 1.7 $ \\
\hline
$\bsb \to K^{+*} \rho^-$      &I  &$12.4$ &$14.1$&$17.5 $&$12.5$ &$14.1$&$17.5$&$0.0$ &$ 0.0$&$0.0 $ \\
$\bsb \to K^{0*} \rho^0$      &II &$0.29$ &$0.03$&$0.94 $&$0.29$ &$0.03$&$0.94$&$0.0$ &$ 0.5$&$0.0 $ \\
$\bsb \to K^{0*} \omega$    &II,VI&$0.24$ &$0.03$&$0.38 $&$0.24$ &$0.04$&$0.38$&$0.6$ &$ 1.3$&$0.0 $ \\
$\bsb \to K^{+*} K^{-*}$      &IV &$2.71$ &$3.02$&$3.66 $&$2.78$ &$3.08$&$3.73$&$2.1$ &$ 1.9$&$1.5 $ \\
$\bsb \to \rho^0 \phi  $      &V  &$0.15$ &$0.21$&$0.45 $&$0.16$ &$0.22$&$0.47$&$5.4$ &$ 5.6$&$4.6 $ \\
$\bsb \to \omega \phi  $      &V  &$0.79$ &$0.01$&$2.40 $&$0.80$ &$0.01$&$2.46$&$0.7$ &$-7.4$&$2.0 $ \\
$\bsb \to K^{0*}\bar{K}^{0*}$ &IV &$2.13$ &$2.70$&$4.06 $&$2.17$ &$2.77$&$4.17$&$2.0$ &$ 2.1$&$2.3 $ \\
$\bsb \to K^{0*}\phi   $      &VI &$0.03$ &$0.12$&$0.48 $&$0.03$ &$0.12$&$0.49$&$3.8$ &$ 2.9$&$2.4 $ \\
$\bsb \to \phi  \phi   $      &VI &$17.4$ &$8.95$&$0.42 $&$17.7$ &$9.08$&$0.42$&$1.1$ &$ 1.1$&$0.5 $ \\
\end{tabular}\end{center}
\end{table}

\begin{table}[tb]
\begin{center}
\caption{ ${\cal B}(B_s\to h_1 h_2)$ (in units of $10^{-6}$) in model II,
with $k^2=m_b^2/2$, $\rho=0.16$, $\eta=0.34$,
$\mhp=200$GeV, $\tan{\beta}=2$, and $\nceff=2,\; 3,\; \infty$. }
\label{brbsm2}
\vspace{0.2cm}
\begin{tabular}{ll rrr rrr rrr}
& & \multicolumn{3}{c}{SM: $\ \ \ {\cal B}$ }
& \multicolumn{6}{c}{Model II: $\ \ \ {\cal B}$ and $\delta {\cal B}[\%]$ } \\
Channel                   & Class &$2$ & $3$& $\infty$&$2$ & $3$& $\infty$ &$2$ & $3$& $\infty$   \\ \hline
$\bsb \to K^+ \pi^-$          & I &$6.33$ &$7.13$&$8.88$& $6.29$ &$7.09$&$8.84$& $-0.6$ &$ -0.6$&$-0.6$ \\
$\bsb \to K^0 \pi^0$          & II&$0.19$ &$0.08$&$0.56$& $0.18$ &$0.07$&$0.55$& $-4.7$ &$ -13.4$&$-2.9$\\
$\bsb \to K^0 \eta $          & VI&$0.34$ &$0.31$&$0.78$& $0.31$ &$0.28$&$0.74$& $-7.5$ &$ -9.8$&$-5.2$ \\
$\bsb \to K^0 \etap $         & VI&$0.57$ &$0.51$&$0.76$& $0.52$ &$0.45$&$0.70$& $-9.5$ &$ -11.6$&$-9.1$\\
$\bsb \to K^+   K^-$          & IV&$10.6$ &$11.7$&$14.0$& $9.58$ &$10.5$&$12.5$& $-10.2$ &$ -10.4$&$-10.8$ \\
$\bsb \to \pi^0 \eta$         & V &$0.04$ &$0.06$&$0.11$& $0.05$ &$0.06$&$0.11$& $ 10.5$ &$ 10.0$&$ 7.3$ \\
$\bsb \to \pi^0 \etap$        & V &$0.04$ &$0.05$&$0.10$& $0.05$ &$0.06$&$0.11$& $ 10.5$ &$ 10.0$&$ 7.3$ \\
$\bsb \to \eta \etap$         & VI&$13.7$ &$15.8$&$20.5$& $12.1$ &$14.0$&$18.2$& $-11.7$ &$ -11.6$&$-11.4$ \\
$\bsb \to \etap\etap$         & VI&$6.77$ &$7.48$&$9.05$& $5.94$ &$6.56$&$7.91$& $-12.5$ &$ -12.7$&$-12.9$ \\
$\bsb \to \eta \eta$          & VI&$6.95$ &$8.35$&$11.5$& $6.21$ &$7.49$&$10.4$& $-10.9$ &$ -10.6$&$-9.9$ \\
$\bsb \to K^0 \bar{K}^0$      & IV&$11.4$ &$13.2$&$17.2$& $10.3$ &$11.9$&$15.6$& $-11.9$ &$ -11.7$&$-11.4$ \\
\hline
$\bsb \to K^{*+} \pi^-$       &I  &$4.04$ &$4.56$&$5.70 $&$4.04$ &$4.56$&$5.70$&$ 0.0$ &$ 0.0$&$ 0.0 $ \\
$\bsb \to K^{+} \rho^-$       &I  &$14.7$ &$16.6$&$20.8 $&$14.7$ &$16.6$&$20.8$&$-0.2$ &$-0.2$&$-0.2 $ \\
$\bsb \to K^{0*} \pi^0$       &II &$0.10$ &$0.003$&$0.29$&$0.10$ &$0.003$&$0.29$&$ 0.8$ &$ 16.5$&$ 0.0 $ \\
$\bsb \to K^{0} \rho^0$      &II  &$0.35$ &$0.04$&$1.11 $&$0.35$ &$0.03$&$1.10$&$-1.1$ &$-16.7$&$-1.0 $ \\
$\bsb \to K^{0} \omega$    &II,VI &$1.14$ &$0.16$&$1.81 $&$1.09$ &$0.15$&$1.80$&$-4.5$ &$-10.3$&$-0.2 $ \\
$\bsb \to \bar{K}^{0*} \eta $    &II,VI &$0.16$ &$0.13$&$0.44 $&$0.15$ &$0.12$&$0.42$&$-7.3$ &$-11.1$&$-4.7 $ \\
$\bsb \to \bar{K}^{0*} \etap$    &II,VI &$0.08$ &$0.02$&$0.16 $&$0.07$ &$0.02$&$0.16$&$-4.7$ &$-17.9$&$-2.9 $ \\
$\bsb \to K^{+} K^{-*}$       &IV &$3.04$ &$3.38$&$4.11 $&$2.74$ &$3.03$&$3.66$&$-10.3$ &$-10.7$&$-11.2 $ \\
$\bsb \to K^{+*}K^{-} $       &IV &$0.89$ &$0.97$&$1.15 $&$0.87$ &$0.95$&$1.14$&$-2.0$ &$-1.6$&$-0.8 $ \\
$\bsb \to \rho \eta   $       &V  &$0.08$ &$0.11$&$0.25 $&$0.09$ &$0.12$&$0.26$&$ 5.4$ &$ 5.6$&$ 4.6 $ \\
$\bsb \to \rho \etap  $       &V  &$0.08$ &$0.11$&$0.24 $&$0.09$ &$0.12$&$0.25$&$ 5.4$ &$ 5.6$&$ 4.6 $ \\
$\bsb \to \omega \eta $       &V  &$0.85$ &$0.01$&$2.59 $&$0.77$ &$0.01$&$2.33$&$-9.9$ &$-7.4$&$-10.3 $ \\
$\bsb \to \omega \etap$       &V  &$0.84$ &$0.01$&$2.55 $&$0.76$ &$0.01$&$2.30$&$-9.9$ &$-7.4$&$-10.3 $ \\
$\bsb \to \pi^0 \phi  $       &V  &$0.13$ &$0.17$&$0.32 $&$0.14$ &$0.18$&$0.34$&$ 10.5$ &$10.0$&$ 7.3 $ \\
$\bsb \to \phi \eta   $       &VI &$1.84$ &$0.75$&$0.07 $&$1.51$ &$0.60$&$0.09$&$-18.3$ &$-20.5$&$ 26.8 $ \\
$\bsb \to \phi \etap  $       &VI &$0.69$ &$0.20$&$1.49 $&$0.55$ &$0.21$&$1.60$&$-21.3$ &$ 7.5$&$ 7.6 $ \\
$\bsb \to K^0 \bar{K}^{0*}$   &IV &$3.22$ &$4.09$&$6.15 $&$2.85$ &$3.64$&$5.54$&$-12.0$ &$-11.3$&$-10.1 $ \\
$\bsb \to K^{0*}\bar{K}^0$    &IV &$0.39$ &$0.31$&$0.18 $&$0.39$ &$0.31$&$0.18$&$-0.6$ &$-1.3$&$-3.2 $ \\
$\bsb \to K^0 \phi    $       &VI &$0.001$&$0.03$&$0.30 $&$0.002$&$0.03$&$0.28$&$ 6.6$ &$ 2.0$&$-6.6 $ \\
\hline
$\bsb \to K^{+*} \rho^-$      &I  &$12.4$ &$14.1$&$17.5 $&$12.4$ &$14.0$&$17.5$&$-0.2$ &$-0.2$&$-0.2 $ \\
$\bsb \to K^{0*} \rho^0$      &II &$0.29$ &$0.03$&$0.94 $&$0.29$ &$0.03$&$0.93$&$-1.1$ &$-16.7$&$-1.0 $ \\
$\bsb \to K^{0*} \omega$    &II,VI&$0.24$ &$0.03$&$0.38 $&$0.23$ &$0.03$&$0.38$&$-4.5$ &$-10.3$&$-0.2 $ \\
$\bsb \to K^{+*} K^{-*}$      &IV &$2.71$ &$3.02$&$3.66 $&$2.43$ &$2.70$&$3.27$&$-10.3$ &$-10.7$&$-11.2 $ \\
$\bsb \to \rho^0 \phi  $      &V  &$0.15$ &$0.21$&$0.45 $&$0.16$ &$0.22$&$0.47$&$ 5.4$ &$ 5.6$&$ 4.6 $ \\
$\bsb \to \omega \phi  $      &V  &$0.79$ &$0.01$&$2.40 $&$0.71$ &$0.01$&$2.16$&$-10.0$ &$-7.4$&$-10.3 $ \\
$\bsb \to K^{0*}\bar{K}^{0*}$ &IV &$2.13$ &$2.70$&$4.06 $&$1.88$ &$2.41$&$3.66$&$-12.1$ &$-11.3$&$-10.2 $ \\
$\bsb \to K^{0*}\phi   $      &VI &$0.03$ &$0.12$&$0.48 $&$0.03$ &$0.11$&$0.44$&$-10.0$ &$-9.2$&$-8.8 $ \\
$\bsb \to \phi  \phi   $      &VI &$17.4$ &$8.95$&$0.42 $&$15.2$ &$7.75$&$0.33$&$-13.0$ &$-13.8$&$-21.4 $ \\
\end{tabular}\end{center}
\end{table}

\begin{table}%[tb]
\begin{center}
\caption{CP-violating asymmetries $\acp (B_s \to h_1 h_2)$
(in percent) in the SM and model III,  with $k^2=m_b^2/2$,
$\rho=0.16$, $\eta=0.34$, $\mhp=200$GeV, $\theta=0^\circ, 30^0$,
and $\nceff=2,\; 3,\; \infty$. }
\label{acpsm1}
\vspace{0.2cm}
\begin{tabular} {ll rrr rrr rrr}
& &\multicolumn{3}{c}{SM }
& \multicolumn{3}{c}{Model III: $\theta=0^\circ,$}& \multicolumn{3}{c}{Model III: $\theta=30^\circ,$} \\
Channel &CP-Class          &$2$ & $3$& $\infty$&$2$ & $3$& $\infty$ &$2$ & $3$& $\infty$   \\ \hline
$\bsb \to K^+ \pi^-$         &$1$&$ 10.2$ &$ 10.2$ &$ 10.3$ &$ 9.95$ &$ 9.96$ &$ 9.97$ &$ 10.2$ &$ 10.2$ &$ 10.3$\\
$\bsb \to K_S^0 \pi^0$       &$2$&$-1.99$ &$-3.98$ &$ 4.50$ &$-3.18$ &$-3.87$ &$ 4.48$ &$-2.60$ &$-4.51$ &$ 4.36$\\
$\bsb \to K_S^0 \eta $       &$2$&$-4.73$ &$-3.62$ &$ 2.93$ &$-4.94$ &$-3.59$ &$ 2.17$ &$-4.76$ &$-4.19$ &$ 1.67$\\
$\bsb \to K_S^0 \etap $      &$2$&$ 0.31$ &$-2.86$ &$-4.74$ &$-0.50$ &$-2.99$ &$-4.96$ &$-1.44$ &$-3.84$ &$-4.77$\\
$\bsb \to K^+   K^-$         &$2$&$-1.71$ &$-1.74$ &$-1.77$ &$-1.40$ &$-1.41$ &$-1.43$ &$-2.38$ &$-2.40$ &$-2.45$\\
$\bsb \to \pi^0 \eta$        &$2$&$-2.72$ &$-0.24$ &$ 2.93$ &$-2.69$ &$-0.24$ &$ 2.92$ &$-2.69$ &$-0.24$ &$ 2.92$\\
$\bsb \to \pi^0 \etap$       &$2$&$-2.72$ &$-0.24$ &$ 2.93$ &$-2.69$ &$-0.24$ &$ 2.92$ &$-2.69$ &$-0.24$ &$ 2.92$\\
$\bsb \to \eta \etap$        &$2$&$ 0.07$ &$ 0.05$ &$ 0.01$ &$ 0.05$ &$ 0.03$ &$ 0.01$ &$-1.08$ &$-1.09$ &$-1.11$\\
$\bsb \to \etap\etap$        &$2$&$-0.16$ &$ 0.03$ &$ 0.36$ &$-0.13$ &$ 0.02$ &$ 0.27$ &$-1.34$ &$-1.21$ &$-0.98$\\
$\bsb \to \eta \eta$         &$2$&$ 0.30$ &$ 0.06$ &$-0.31$ &$ 0.24$ &$ 0.05$ &$-0.25$ &$-0.81$ &$-0.97$ &$-1.24$\\
$\bsb \to K^0 \bar{K}^0$     &$2$&$ 0.05$ &$ 0.04$ &$ 0.04$ &$ 0.03$ &$ 0.03$ &$ 0.03$ &$-1.19$ &$-1.18$ &$-1.17$\\
\hline
$\bsb \to K^{*+} \pi^-$      &$1$&$ 0.34$ &$ 0.34$ &$ 0.34$ &$ 0.34$ &$ 0.34$ &$ 0.34$ &$ 0.34$ &$ 0.34$ &$ 0.34$\\
$\bsb \to K^{+} \rho^-$      &$1$&$ 5.56$ &$ 5.56$ &$ 5.55$ &$ 5.52$ &$ 5.51$ &$ 5.51$ &$ 5.60$ &$ 5.60$ &$ 5.59$\\
$\bsb \to K^{0*} \pi^0$      &$1$&$-8.45$ &$-26.6$ &$ 6.17$ &$-8.59$ &$-41.2$ &$ 6.17$ &$-8.40$ &$-40.1$ &$ 6.27$\\
$\bsb \to K_S^{0} \rho^0$    &$1$&$-22.6$ &$-16.3$ &$ 15.5$ &$-21.1$ &$-7.99$ &$ 14.8$ &$-20.3$ &$-16.6$ &$ 15.4$\\
$\bsb \to K_S^{0} \omega$    &$2$&$ 3.88$ &$-1.60$ &$ 4.53$ &$ 3.15$ &$-2.01$ &$ 4.65$ &$ 2.78$ &$-2.90$ &$ 4.64$\\
$\bsb \to \bar{K}^{0*} \eta $&$1$&$-27.3$ &$-0.36$ &$ 23.1$ &$-19.5$ &$-0.01$ &$ 18.6$ &$-21.6$ &$-4.60$ &$ 18.7$\\
$\bsb \to \bar{K}^{0*} \etap$&$1$&$ 52.8$ &$ 30.8$ &$-45.0$ &$ 39.9$ &$ 14.1$ &$-36.7$ &$ 42.5$ &$-2.09$ &$-35.7$\\
$\bsb \to K^{+} K^{-*}$      &$1$&$-30.5$ &$-30.9$ &$-31.7$ &$-18.8$ &$-19.0$ &$-19.4$ &$-23.9$ &$-24.2$ &$-24.6$\\
$\bsb \to K^{+*}K^{-} $      &$1$&$ 1.40$ &$ 1.45$ &$ 1.53$ &$ 1.37$ &$ 1.42$ &$ 1.49$ &$ 1.37$ &$ 1.41$ &$ 1.49$\\
$\bsb \to \rho \eta   $      &$2$&$-3.03$ &$-0.26$ &$ 3.02$ &$-3.02$ &$-0.26$ &$ 3.01$ &$-3.02$ &$-0.26$ &$ 3.01$\\
$\bsb \to \rho \etap  $      &$2$&$-3.03$ &$-0.26$ &$ 3.02$ &$-3.02$ &$-0.26$ &$ 3.01$ &$-3.02$ &$-0.26$ &$ 3.01$\\
$\bsb \to \omega \eta $      &$2$&$-0.87$ &$-1.01$ &$-0.84$ &$-0.71$ &$-1.02$ &$-0.68$ &$-1.67$ &$-1.02$ &$-1.74$\\
$\bsb \to \omega \etap$      &$2$&$-0.87$ &$-1.01$ &$-0.84$ &$-0.71$ &$-1.02$ &$-0.68$ &$-1.67$ &$-1.02$ &$-1.74$\\
$\bsb \to \pi^0 \phi  $      &$2$&$-2.72$ &$-0.24$ &$ 2.93$ &$-2.69$ &$-0.24$ &$ 2.92$ &$-2.69$ &$-0.24$ &$ 2.92$\\
$\bsb \to \phi \eta   $      &$2$&$ 0.49$ &$ 0.17$ &$ 2.87$ &$-0.71$ &$-1.02$ &$-0.68$ &$-1.67$ &$-1.02$ &$-1.74$\\
$\bsb \to \phi \etap  $      &$2$&$-0.31$ &$ 0.25$ &$-0.74$ &$-0.71$ &$-1.02$ &$-0.68$ &$-1.67$ &$-1.02$ &$-1.74$\\
$\bsb \to K_S^0\bar{K}^{0*}$ &$1$&$-1.25$ &$-1.21$ &$-1.13$ &$-1.02$ &$-0.98$ &$-0.93$ &$-6.98$ &$-6.46$ &$-5.72$\\
$\bsb \to K^{0*} K_S^0$      &$1$&$-0.02$ &$-0.03$ &$-0.04$ &$-0.02$ &$-0.03$ &$-0.04$ &$-0.02$ &$-0.03$ &$-0.04$\\
$\bsb \to K_S^0 \phi   $     &$2$&$-3.38$ &$-3.45$ &$-3.27$ &$-2.88$ &$-3.45$ &$-3.30$ &$-4.42$ &$-3.47$ &$-3.84$\\
\hline
$\bsb \to K^{+*} \rho^-$     &$1$&$ 5.56$ &$ 5.56$ &$ 5.55$ &$ 5.52$ &$ 5.51$ &$ 5.51$ &$ 5.60$ &$ 5.60$ &$ 5.59$\\
$\bsb \to K^{0*} \rho^0$     &$1$&$-22.6$ &$-16.3$ &$ 15.5$ &$-21.1$ &$-7.99$ &$ 14.8$ &$-20.3$ &$-16.6$ &$ 15.4$\\
$\bsb \to K^{0*} \omega$     &$1$&$ 25.5$ &$ 13.1$ &$ 5.43$ &$ 20.6$ &$ 8.57$ &$ 5.37$ &$ 20.9$ &$ 4.39$ &$ 5.46$\\
$\bsb \to K^{+*} K^{-*}$     &$3$&$-3.86$ &$-3.87$ &$-3.90$ &$-3.25$ &$-3.27$ &$-3.29$ &$-4.05$ &$-4.07$ &$-4.10$\\
$\bsb \to \rho^0 \phi  $     &$3$&$-3.03$ &$-0.26$ &$ 3.02$ &$-3.02$ &$-0.26$ &$ 3.01$ &$-3.02$ &$-0.26$ &$ 3.01$\\
$\bsb \to \omega \phi  $     &$3$&$-0.87$ &$-1.01$ &$-0.84$ &$-0.71$ &$-1.02$ &$-0.68$ &$-1.67$ &$-1.02$ &$-1.74$\\
$\bsb\to K^{0*}\bar{K}^{0*}$ &$3$&$ 0.05$ &$ 0.05$ &$ 0.04$ &$ 0.04$ &$ 0.03$ &$ 0.03$ &$-1.18$ &$-1.13$ &$-1.06$\\
$\bsb \to K^{0*}\phi   $     &$1$&$ 4.41$ &$ 3.46$ &$ 3.01$ &$ 2.90$ &$ 2.40$ &$ 2.15$ &$-3.03$ &$-2.27$ &$-1.91$\\
$\bsb \to \phi  \phi   $     &$3$&$ 0.05$ &$ 0.06$ &$ 0.12$ &$ 0.04$ &$ 0.04$ &$ 0.06$ &$-1.19$ &$-1.24$ &$-1.79$\\
\end{tabular}\end{center}
\end{table}

\begin{table}%[tb]
\begin{center}
\caption{$\acp(B_s \to h_1 h_2 )$ (in percent) in models I and II,
with $k^2=m_b^2/2$, $\rho=0.16$, $\eta=0.34$,
$\mhp=200$GeV, $\tan{\beta}=2$, and $\nceff=2,\; 3,\; \infty$. }
\label{acpm1}
\begin{tabular} {ll rrr rrr rrr}
& & \multicolumn{3}{c}{ SM }
& \multicolumn{3}{c}{Model I} & \multicolumn{3}{c}{Model II} \\
Channel& CP-class          &$2$ & $3$& $\infty$ &$2$ & $3$& $\infty$ &$2$ & $3$& $\infty$   \\ \hline
$\bsb \to K^+ \pi^-$         &$1$&$ 10.2$ &$ 10.2$ &$ 10.3$ &$ 10.2$ &$ 10.3$ &$ 10.3$ &$ 10.3$ &$ 10.3$ &$ 10.3$  \\
$\bsb \to K_S^0 \pi^0$       &$2$&$-1.99$ &$-3.98$ &$ 4.50$ &$-1.98$ &$-3.98$ &$ 4.79$ &$-1.63$ &$-4.01$ &$ 4.84$   \\
$\bsb \to K_S^0 \eta $       &$2$&$-4.73$ &$-3.62$ &$ 2.93$ &$-4.74$ &$-3.62$ &$ 2.92$ &$-4.65$ &$-3.63$ &$ 3.11$  \\
$\bsb \to K_S^0 \etap $      &$2$&$ 0.31$ &$-2.86$ &$-4.74$ &$ 0.28$ &$-2.86$ &$-4.75$ &$ 0.52$ &$-2.82$ &$-4.64$  \\
$\bsb \to K^+   K^-$         &$2$&$-1.71$ &$-1.74$ &$-1.77$ &$-1.70$ &$-1.72$ &$-1.76$ &$-1.80$ &$-1.82$ &$-1.86$  \\
$\bsb \to \pi^0 \eta$        &$2$&$-2.72$ &$-0.24$ &$ 2.93$ &$-2.59$ &$-0.23$ &$ 2.84$ &$-2.59$ &$-0.23$ &$ 2.84$  \\
$\bsb \to \pi^0 \etap$       &$2$&$-2.72$ &$-0.24$ &$ 2.93$ &$-2.59$ &$-0.23$ &$ 2.84$ &$-2.59$ &$-0.23$ &$ 2.84$  \\
$\bsb \to \eta \etap$        &$2$&$ 0.07$ &$ 0.05$ &$ 0.01$ &$ 0.07$ &$ 0.05$ &$ 0.01$ &$ 0.07$ &$ 0.05$ &$ 0.01$  \\
$\bsb \to \etap\etap$        &$2$&$-0.16$ &$ 0.03$ &$ 0.36$ &$-0.16$ &$ 0.03$ &$ 0.36$ &$-0.16$ &$ 0.04$ &$ 0.39$  \\
$\bsb \to \eta \eta$         &$2$&$ 0.30$ &$ 0.06$ &$-0.31$ &$ 0.30$ &$ 0.06$ &$-0.31$ &$ 0.31$ &$ 0.06$ &$-0.32$  \\
$\bsb \to K^0 \bar{K}^0$     &$2$&$ 0.05$ &$ 0.04$ &$ 0.04$ &$ 0.04$ &$ 0.04$ &$ 0.04$ &$ 0.05$ &$ 0.05$ &$ 0.05$ \\
\hline
$\bsb \to K^{*+} \pi^-$      &$1$&$ 0.34$ &$ 0.34$ &$ 0.34$ &$ 0.34$ &$ 0.34$ &$ 0.34$&$ 0.34$ &$ 0.34$ &$ 0.34$\\
$\bsb \to K^{+} \rho^-$      &$1$&$ 5.56$ &$ 5.56$ &$ 5.55$ &$ 5.57$ &$ 5.57$ &$ 5.56$&$ 5.58$ &$ 5.57$ &$ 5.57$\\
$\bsb \to K^{0*} \pi^0$      &$1$&$-8.45$ &$-26.6$ &$ 6.17$ &$-8.44$ &$-25.3$ &$ 6.18$&$-8.40$ &$-22.9$ &$ 6.18$\\
$\bsb \to K_S^{0} \rho^0$    &$1$&$-22.6$ &$-16.3$ &$ 15.5$ &$-22.6$ &$-16.2$ &$ 15.5$&$-22.9$ &$-19.8$ &$ 15.7$\\
$\bsb \to K_S^{0} \omega$    &$2$&$ 3.88$ &$-1.60$ &$ 4.53$ &$ 3.87$ &$-1.62$ &$ 4.53$&$ 4.03$ &$-1.49$ &$ 4.50$\\
$\bsb \to \bar{K}^{0*} \eta $&$1$&$-27.3$ &$-0.36$ &$ 23.1$ &$-27.2$ &$-0.36$ &$ 23.1$&$-29.5$ &$-0.50$ &$ 24.3$ \\
$\bsb \to \bar{K}^{0*} \etap$&$1$&$ 52.8$ &$ 30.8$ &$-45.0$ &$ 52.6$ &$ 30.0$ &$-44.8$&$ 55.4$ &$ 37.1$ &$-46.5$\\
$\bsb \to K^{+} K^{-*}$      &$1$&$-30.5$ &$-30.9$ &$-31.7$ &$-29.9$ &$-30.4$ &$-31.3$&$-33.9$ &$-34.5$ &$-35.6$\\
$\bsb \to K^{+*}K^{-} $      &$1$&$ 1.40$ &$ 1.45$ &$ 1.53$ &$ 1.43$ &$ 1.47$ &$ 1.53$&$ 1.44$ &$ 1.47$ &$ 1.54$\\
$\bsb \to \rho \eta   $      &$2$&$-3.03$ &$-0.26$ &$ 3.02$ &$-2.96$ &$-0.25$ &$ 2.96$&$-2.96$ &$-0.25$ &$ 2.96$\\
$\bsb \to \rho \etap  $      &$2$&$-3.03$ &$-0.26$ &$ 3.02$ &$-2.96$ &$-0.25$ &$ 2.96$&$-2.96$ &$-0.25$ &$ 2.96$\\
$\bsb \to \omega \eta $      &$2$&$-0.87$ &$-1.01$ &$-0.84$ &$-0.87$ &$-1.05$ &$-0.84$&$-0.91$ &$-1.05$ &$-0.89$\\
$\bsb \to \omega \etap$      &$2$&$-0.87$ &$-1.01$ &$-0.84$ &$-0.87$ &$-1.05$ &$-0.84$&$-0.91$ &$-1.05$ &$-0.89$\\
$\bsb \to \pi^0 \phi  $      &$2$&$-2.72$ &$-0.24$ &$ 2.93$ &$-2.59$ &$-0.23$ &$ 2.84$&$-2.59$ &$-0.23$ &$ 2.84$\\
$\bsb \to \phi \eta   $      &$2$&$ 0.49$ &$ 0.17$ &$ 2.87$ &$-0.87$ &$-1.05$ &$-0.84$&$-0.91$ &$-1.05$ &$-0.89$\\
$\bsb \to \phi \etap  $      &$2$&$-0.31$ &$ 0.25$ &$-0.74$ &$-0.87$ &$-1.05$ &$-0.84$&$-0.91$ &$-1.05$ &$-0.89$\\
$\bsb \to K_S^0\bar{K}^{0*}$ &$1$&$-1.25$ &$-1.21$ &$-1.13$ &$-1.24$ &$-1.20$ &$-1.12$&$-1.31$ &$-1.26$ &$-1.18$\\
$\bsb \to K^{0*} K_S^0$      &$1$&$-0.02$ &$-0.03$ &$-0.04$ &$-0.02$ &$-0.03$ &$-0.04$&$-0.02$ &$-0.03$ &$-0.04$ \\
$\bsb \to K_S^0 \phi   $     &$2$&$-3.38$ &$-3.45$ &$-3.27$ &$-3.39$ &$-3.45$ &$-3.27$&$-3.67$ &$-3.45$ &$-3.26$ \\
\hline
$\bsb \to K^{+*} \rho^-$     &$1$&$ 5.56$ &$ 5.56$ &$ 5.55$ &$ 5.57$ &$ 5.57$ &$ 5.56$&$ 5.58$ &$ 5.57$ &$ 5.57$ \\
$\bsb \to K^{0*} \rho^0$     &$1$&$-22.6$ &$-16.3$ &$ 15.5$ &$-22.6$ &$-16.2$ &$ 15.5$&$-22.9$ &$-19.8$ &$ 15.7$\\
$\bsb \to K^{0*} \omega$     &$1$&$ 25.5$ &$ 13.1$ &$ 5.43$ &$ 25.4$ &$ 12.9$ &$ 5.44$&$ 26.7$ &$ 14.5$ &$ 5.45$\\
$\bsb \to K^{+*} K^{-*}$     &$3$&$-3.86$ &$-3.87$ &$-3.90$ &$-3.83$ &$-3.85$ &$-3.89$&$-3.97$ &$-3.99$ &$-4.02$\\
$\bsb \to \rho^0 \phi  $     &$3$&$-3.03$ &$-0.26$ &$ 3.02$ &$-2.96$ &$-0.25$ &$ 2.96$&$-2.96$ &$-0.25$ &$ 2.96$ \\
$\bsb \to \omega \phi  $     &$3$&$-0.87$ &$-1.01$ &$-0.84$ &$-0.87$ &$-1.05$ &$-0.84$&$-0.91$ &$-1.05$ &$-0.89$ \\
$\bsb\to K^{0*}\bar{K}^{0*}$ &$3$&$ 0.05$ &$ 0.05$ &$ 0.04$ &$ 0.05$ &$ 0.05$ &$ 0.04$&$ 0.06$ &$ 0.05$ &$ 0.05$ \\
$\bsb \to K^{0*}\phi   $     &$1$&$ 4.41$ &$ 3.46$ &$ 3.01$ &$ 4.28$ &$ 3.38$ &$ 2.96$&$ 4.81$ &$ 3.74$ &$ 3.25$ \\
$\bsb \to \phi  \phi   $     &$3$&$ 0.05$ &$ 0.06$ &$ 0.12$ &$ 0.05$ &$ 0.05$ &$ 0.12$&$ 0.06$ &$ 0.06$ &$ 0.14$ \\
\end{tabular}\end{center}
\end{table}

\newpage

\listoffigures

\newpage

\newpage

\newpage

\newpage

\newpage

\newpage

\newpage

\newpage

\end{document}